\documentclass[english,final,superscriptaddress]{iopart}
\usepackage[T1]{fontenc}
\usepackage[latin9]{inputenc}
\usepackage{amsmath}
\usepackage{amssymb}
\usepackage{graphicx}
\usepackage{esint}

\makeatletter
\usepackage{iopams}
\usepackage{setstack}

\makeatother

\usepackage{babel}
\begin{document}

\title{Thermal Baths as Quantum Resources: More Friends than Foes?}

\author{Gershon Kurizki$^{1}$, Ephraim Shahmoon$^{2}$ and Analia Zwick$^{1}$}

\address{$^{1}$Weizmann Institute of Science, Rehovot 76100, Israel}

\address{$^{2}$Department of Physics, Harvard University, Cambridge MA 02138,
USA}
\begin{abstract}
In this article we argue that thermal reservoirs (baths) are potentially
useful resources in processes involving atoms interacting with quantized
electromagnetic fields and their applications to quantum technologies.
One may try to suppress the bath effects by means of dynamical control,
but such control does not always yield the desired results. We wish
instead to take advantage of bath effects, that do not obliterate
\textquotedblleft quantumness\textquotedblright{} in the system-bath
compound. To this end, three possible approaches have been pursued
by us: \emph{(i) }Control of a quantum system faster than the correlation
time of the bath to which it couples: Such control allows us to reveal
quasi-reversible/coherent dynamical phenomena of quantum open systems,
manifest by the quantum Zeno or anti-Zeno effects (QZE or AZE, respectively).
Dynamical control methods based on the QZE are aimed not only at protecting
the quantumness of the system, but also diagnosing the bath spectra
or transferring quantum information via noisy media. By contrast,
AZE-based control is useful for fast cooling of thermalized quantum
systems. \emph{(ii)} Engineering the coupling of quantum systems to
selected bath modes: This approach, based on field -atom coupling
control in cavities, waveguides and photonic band structures, allows
to drastically enhance the strength and range of atom-atom coupling
through the mediation of the selected bath modes. More dramatically,
it allows us to achieve \emph{bath-induced entanglement} that may
appear paradoxical if one takes the conventional view that coupling
to baths destroys quantumness. \emph{(iii)} Engineering baths with
appropriate non-flat spectra: This approach is a prerequisite for
the construction of the simplest and most efficient quantum heat machines
(engines and refrigerators). We may thus conclude that often thermal
baths are ``more friends than foes'' in quantum technologies.
\end{abstract}
\maketitle

\section{Introduction}

In this article we wish to \emph{\textquotedblleft make out a case\textquotedblright{}}
for thermal reservoirs (baths) as potentially useful resources in
quantum optics \cite{scully1997quantum}, namely, in processes involving
matter interacting with quantized electromagnetic fields, and their
applications to quantum technologies: quantum information processing
\cite{nie00a,Gisin_Quantum_cryptog_2002,Sergienko_2005_QCC}, quantum
sensing and metrology \cite{giovannetti_2011_advances,wolfgramm_2013_entanglement,schmidt_2005_spectroscopy,hempel_2013_entanglement,ockeloen_2013_quantum},
as well as quantum thermodynamics \cite{bookMahler,scully2003extracting,scully2011quantum}.
In general, there is little we can do to avoid the ubiquitous presence
of environments described as thermal baths in contact with quantum
systems: with very few exceptions, all quantum systems are open \cite{bookBreuer,RevModPhys75715}.
One may try to reduce the bath effects on the quantum system of interest
by means of dynamical control, originally developed to suppress bath-induced
decoherence or dissipation \cite{vio98,aga01,aga00b,shi04,vit01a,vio03,kho05}.
Yet such control does not always yield the desired results, hence
we wish to advocate a different strategy that may be colloquially
summarized as follows: \emph{\textquotedblleft If you can\textquoteright t
fight the bath \textendash{} join it\textquotedblright }, namely,
take advantage of its effects, particularly those that do not obliterate
\textquotedblleft quantumness\textquotedblright{} in the system-bath
compound. To this end, three possible approaches may be pursued, to
be discussed in the subsequent sections \footnote{Cooperative (Dicke) effects mediated by the bath are outside the scope
of this article - cf. the following articles:\cite{dicke1954coherence,prasad2000polarium,PhysRevA.84.023805,scully2006directed,scully2009super,2007JPhB...40..105M}}:
\begin{itemize}
\item \textbf{Control a quantum system }faster than\textbf{ the correlation
(memory) time of the bath to which it couples}: Such control allows
us to reveal quasi-reversible/ coherent dynamical phenomena of quantum
open systems, manifest by the quantum Zeno or anti-Zeno effects (QZE
or AZE, respectively) \cite{mis77,kof00,kof01,facchi2001quantum,lan83,fis01}.
Dynamical control methods based on the QZE are aimed at protecting
the quantumness of the system \cite{kof01,kof04,kof01a,kof96,pel04,gor08PRL,kof05,fis01,gor06a,bar04,gor05a,gor06b,JPB,gor07PRA,fac02,Bri05},
but also diagnosing the bath spectra and transferring quantum information
via noisy media (Sec. \ref{sec: 2_Control-within-the}). By contrast,
AZE-based control is useful for fast cooling of thermalized quantum
systems \cite{ere08Nature,gor10NJP,gor09} (Sec. \ref{sec:5_Thermodynamic-control-via}).
\item \textbf{Engineer the coupling of quantum systems to selected bath
modes}: This approach, based on field -atom coupling control in cavities
\cite{kur96,ShermanPRA92,KozhekinPRA96} and photonic band structures
\cite{she92,kurizki1993quantum,li2008fabrication,pet02,kurizki1994special,friedler2005deterministic},
allows to drastically modify bath-mediated exchange of virtual quanta
between quantum systems and thereby enormously enhance their coupling
\cite{kur90,shahmoon2014giant,shahmoon2014non}. Not less dramatically,
such engineering allows us to achieve bath-induced entanglement \cite{friedler2005long,petrosyan2001photon,shahmoon2011strongly,friedler2004giant,mandilara2007control,shahmoon2013nonradiative,shahmoon2013engineering,shahmoon2013dispersion}
that may appear paradoxical if one takes the conventional view that
coupling to baths destroys quantumness \cite{bookBreuer,RevModPhys75715}
(Sec. \ref{sec: 3_Bath-induced-entanglement}-\ref{sec:4_Long-range-bath-induced-dispersi}). 
\item \textbf{Select or engineer baths with appropriate non-flat spectra}:
This approach is a prerequisite for the construction of the simplest
and most efficient quantum heat machines (engines and refrigerators)
\cite{gelbwaser2013minimal,gelbwaser2013work,kolavr2012quantum} and
for investigating their ability to attain the absolute zero \cite{kolavr2012quantum}
(Sec. \ref{sec:6_Heat-machine-design-by}).
\end{itemize}
Our conclusions and outlook to forthcoming research along the discussed
lines are presented in Sec. \ref{sec:7_Conclusions-and-Outlook}.

\section{Control within the bath memory-time: Zeno \& anti-Zeno dynamics\label{sec: 2_Control-within-the}}

Our theory of quantum systems whose weak interaction with thermal
baths is dynamically controlled \cite{kof01,kof04,gor06a,gor05a,JPB,gordon_scalability_2011,clausen_bath-optimized_2010,clausen_task-optimized_2012}
treats all kinds of such control, be it coherent or projective (non-unitary),
continuous or pulsed, as generalized forms of two generic effects
or control paradigms. One is \emph{
\[
\textrm{Minimized bath effect \ensuremath{\equiv}Quantum Zeno effect (QZE),}
\]
}which \emph{minimizes} (under constraints on the control energy)
the integral product (overlap) of two functions: $G(\omega)$, the
coupling spectrum of the bath (obtained by Fourier-transforming its
autocorrelation function) and a spectral \textquotedblleft filter\textquotedblright{}
function $F_{t}(\omega)$ determined by the control field-intensity
spectrum and its time duration t. It is the \textquotedblleft filter\textquotedblright{}
function that provides the control handle on our ability to optimally
execute a desired task in the presence of a given bath. In the presence
of several baths (a common situation), both $G(\omega)$ and $F_{t}(\omega)$
functionals are represented by matrices \cite{gordon_scalability_2011,clausen_bath-optimized_2010,clausen_task-optimized_2012}.

QZE-based control is required in operational tasks related to quantum
information its storage and transmission \cite{petrosyan_reversible_2009,BenskyQIP11,Zwick_Optimized_2014,escher_optimized_2011,bensky_optimizing_2012},
where bath effects are detrimental and must be suppressed. Regardless
of the chosen form of control, the controlled-system dynamics must
then be \emph{Zeno-like}, namely, result in suppressed system-bath
interaction.

The alternative paradigm is\emph{
\[
\textrm{Maximized bath effect\ensuremath{\equiv}Anti-Zeno effect (AZE),}
\]
}which amounts to \emph{maximized overlap} of $G(\omega)$ and $F_{t}(\omega)$
(under control-energy constraints, as for QZE). AZE-based control
is instrumental for non-unitary operations that entail changes of
the system\textquoteright s entropy. Such operations benefit from
efficient interaction with a bath for their execution. Examples are
measurements used to cool (purify) a quantum system \cite{gor09},
equilibrate (thermalize) it with a bath \cite{gor10NJP,clausen_bath-optimized_2010},
or harvest energy from the bath. If the underlying dynamics is anti-Zeno-like
\cite{kof00,ere08Nature}, system-bath interaction will be enhanced
and thereby facilitate these tasks.

Certain tasks may involve state transfer or entanglement via the bath,
which require maximized bipartite coupling, but minimized single-partite
coupling with the bath \cite{gor06b,gordon_scalability_2011,BarGillPRL11a,durga11}.
For such tasks, a more subtle interplay of Zeno and anti-Zeno dynamics
may be optimal and depend on the quantum statistics of the bath \cite{RaoPRA11}. 

We have therefore developed a general approach that allows to optimize
the interaction of a quantum system with the environment so as to
execute a given operation, be it non-unitary or unitary, such as state-
transfer or storage with maximized fidelity, purification/entropy-minimization,
entanglement distribution, or energy transfer \cite{clausen_bath-optimized_2010,clausen_task-optimized_2012}.
This approach consists in designing the temporal dependence of the
Hamiltonian that governs the system by variational minimization or
maximization (as the case may be) of a state-dependent functional
chosen to quantify the success probability of the operation. To this
end, the temporal control must be faster than the bath correlation
time \cite{gor08PRL,clausen_bath-optimized_2010,clausen_task-optimized_2012}.
This approach not only provides protection from adverse effects of
the bath, namely, quantum-state decoherence, but actually benefits
from the system-bath interactions for the realization of a given non-unitary
task. More formally, it maximizes the fidelity of any given quantum
operation on a multidimensional Hilbert space for the baths or noise
sources at hand. Its main merit is that it is not restricted to pulsed
forms of control, and therefore can drastically reduce the energy
required to execute a task by resorting to a smoothly varying field,
thereby reducing the errors incurred by control \cite{clausen_bath-optimized_2010}.

\subsection{Control for bath diagnostics\label{sub:2.1_Control-for-bath}}

We shall discuss applications of dynamical control of the system-bath
coupling that go beyond its conventional use as a means of fighting
decoherence \cite{vio98,aga01,aga00b,shi04,vit01a,vio03,kho05,kof04,gor08PRL,kof05,gor06a,souza2012robust}.
The first application of such control is as a tool of bath\textendash spectrum
diagnostics. Such diagnostics has the goal of revealing the dynamics
of decoherence processes and their underlying bipartite and multipartite
interactions (collisions).

The basis for this diagnostics is the Kofman-Kurizki (KK) universal
formula \cite{kof00,kof01,kof04}

\begin{equation}
R(t)=\intop F_{t}(\omega)G(\omega)d\omega\:\leftrightarrow\textrm{change}\:\underset{\mathrm{Filter}}{\underbrace{F_{t}(\omega)}}\Rightarrow\textrm{infer}\underset{\!\mathrm{Coupling-spectrum}}{\!\!\!\!\!\!\!\!\!\!\!\!\!\!\!\!\!\!\!\!\!\!\!\underbrace{G(\omega)}}.\label{eq:1_R(t)}
\end{equation}
The diagnostic method consists in changing the filter function $F_{t}(\omega)$,
\emph{e.g.}, by varying the control-field Rabi frequency, recording
the resulting decoherence rate $R(t)$ and deducing $G(\omega)$ from
Eq. (\ref{eq:1_R(t)}). To this end, the system, \emph{e.g.} a qubit,
is initially taken to be in a superposition of its excited ($\left|\uparrow\right\rangle $)
and ground ($\left|\downarrow\right\rangle $) energy states. This
initial superposition state 
\begin{equation}
\left|\psi(0)\right\rangle =\cos\Theta\left|\uparrow\right\rangle +e^{-i\phi}\sin\Theta\left|\downarrow\right\rangle ,\label{eq:init-stat}
\end{equation}
is subject to bath-induced decoherence (pure dephasing). It then has,
at time $t$, a mean coherence that decays in a fashion dependent
on $R(t)$ 
\begin{equation}
\left\langle \sigma_{x}(t)\right\rangle =e^{-R(t)\,t}\left\langle \sigma_{x}(0)\right\rangle ,\label{eq:<sigma_x>-1}
\end{equation}
which is inferred from the probabilities of measuring the system in
the symmetric or antisymmetric superpositions of energy states (Fig.
\ref{fig:1})

\begin{equation}
p_{\pm}(t)=\frac{1\pm e^{-R(t)\,t}}{2}.\label{eq:p+-}
\end{equation}

We have demonstrated (in collaboration with Davidson\textquoteright s
group) \cite{alm11} the ability to infer the bath-coupling spectrum
$G(\omega)$ via formula (\ref{eq:1_R(t)}) by measurements performed
on a large ensemble of cold atoms in an optical trap. A field with
narrow spectral band was used to realize a filter function $F_{t}(\omega)$
that scanned the overlap integral in Eq. (\ref{eq:1_R(t)}) upon varying
the field strength (Rabi frequency). By measuring the decoherence
rate $R(t)$ as a function of the filter value we could infer the
bath-coupling spectrum in the weak-coupling limit. This demonstration
has experimentally established that the Kofman-Kurizki (KK) universal
formula (\ref{eq:1_R(t)}) allows the design of dynamical control
(continuous-wave or pulse sequence) that is optimally adapted to the
measured coupling spectrum of the bath.

\subsection{Maximized information on the bath by dynamical control\label{sub:2.2_Maximized-information-on}}

\begin{figure}
\begin{centering}
\includegraphics[width=0.8\textwidth]{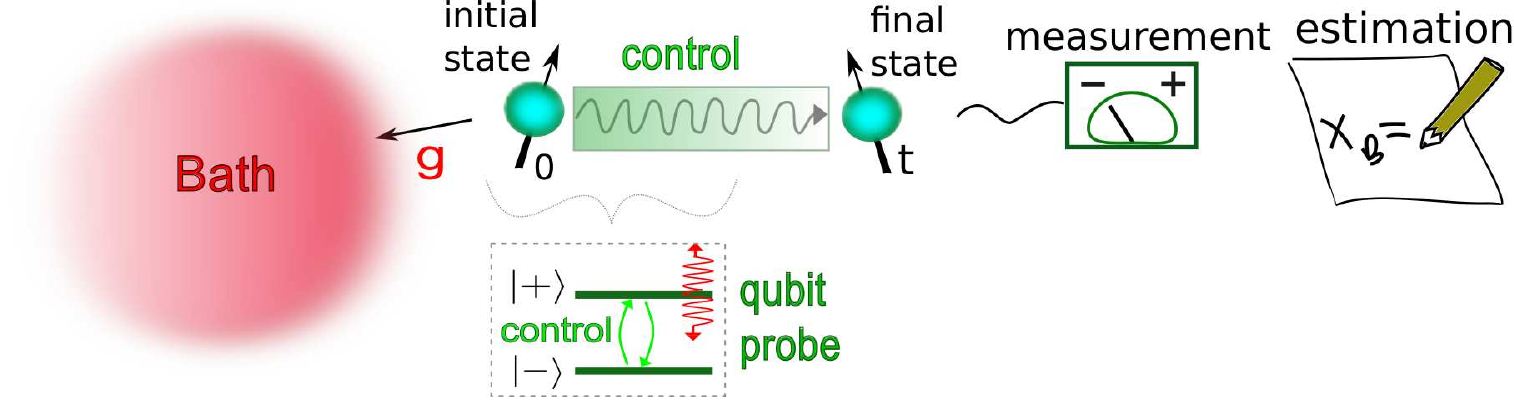}
\par\end{centering}

\protect\caption{\label{fig:1}Schematic view of probing bath parameters by a qubit
that undergoes bath-induced decoherence while being subject to dynamical
control. Information on the bath that yields estimation of its parameters
is extracted from a measurement of the qubit in the $\left|\pm\right\rangle =\frac{1}{\sqrt{2}}(\left|\uparrow\right\rangle \pm\left|\downarrow\right\rangle )$
basis of Eq. (\ref{eq:init-stat}).}
\end{figure}

We have recently been studying the maximum information obtainable
on unknown spectral parameters of a bath (environment) by controlled
spin qubits that serve as its probes \cite{Zwick_NatCom_2015}. This
information is important for maximizing the sensitivity of spin probes
at nanoscales, serving as magnetometers, thermometers, sensors for
imaging or monitoring chemical and biological processes \cite{neumann2013high,mcguinness2011quantum,SmithPNAS12,alvarezcoherent2013}. 

By using tools of quantum estimation theory, we can find the precision
of estimating key parameters of environmental noises (baths) that
the spin (qubit) can probe. These include the probe-bath coupling
strength $g$, the correlation time of generic bath spectra $\tau_{c}$,
as well as their dephasing time $T_{2}$. By optimizing the dynamical
control on the probe under realistic constraints one may achieve the
best accuracy of estimating these parameters by the least number of
measurements possible.

To this end, we minimize the relative error of estimating a bath parameter
$x_{B}$ by means of the dependence of the decoherence rate of a qubit-probe
$R(t)\equiv R(x_{B},t)$ on $G(\omega)\equiv G(x_{B},\omega)$, as
described by Eqs. (\ref{eq:1_R(t)})-(\ref{eq:p+-}) (Fig. \ref{fig:1}).
This error obeys the bound

\begin{equation}
\varepsilon(x_{B},t)=\frac{\delta x_{B}}{x_{B}}\geq\frac{1}{x_{B}\sqrt{N_{m}\mathcal{F_{Q}}(x_{B},t)}}.
\end{equation}
Here we have introduced the number of measurements, $N_{m}$, and
the quantum Fisher Information (QFI) \cite{Caves_1994_fisher,Paris2009_QUANTUM-ESTIMATION,Paris2014_Characterization-of-classical-Gaussian,escher2011general}
for the qubit probe that is subject to dephasing as well as dynamical
control
\begin{equation}
\mathcal{F_{Q}}(x_{B},t)=sin^{2}(2\Theta)\frac{e^{-2R(x_{B},t)\,t}}{1-e^{-2R(x_{B},t)\,t}}\left(\frac{\partial R(x_{B},t)t}{\partial x_{B}}\right)^{2},
\end{equation}
where $\Theta$ is as in Eq. (\ref{eq:init-stat}). In general, we
can minimize the relative error per measurement by maximizing QFI:

\begin{equation}
\mathcal{F_{Q}}(x_{B},t_{opt})=\underset{t}{max}\mathcal{F_{Q}}(x_{B},t)
\end{equation}
which amounts to preparing the optimal initial state (\ref{eq:init-stat})
with $\Theta=\frac{\pi}{4}$, measuring the qubit at the optimal time
and efficiently controlling the quantum probe.

To demonstrate the potential of this approach, we may estimate, for
example, the correlation time $\tau_{c}$, a key parameter of Ornstein-Uhlenbeck
processes characterized by Lorenztian bath spectra,
\begin{equation}
G(x_{B},\omega)=\frac{g^{2}\tau_{c}}{\pi(1+\omega^{2}\tau_{c}^{2})},\label{eq:Gw}
\end{equation}
assuming that the system-bath coupling strength $g$ is known.

\begin{figure}
\begin{centering}
\includegraphics[width=1\textwidth]{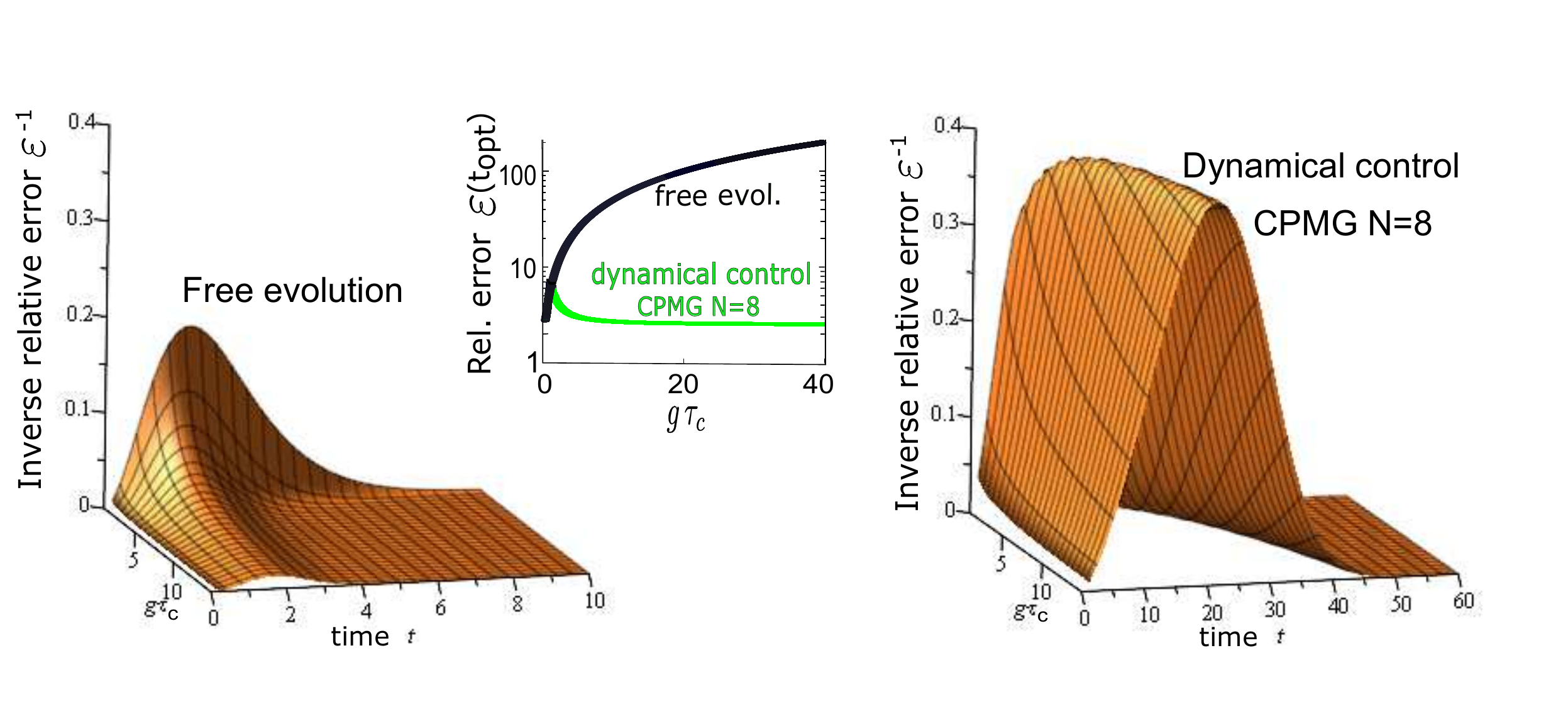}
\par\end{centering}

\protect\caption{\label{fig:2}Estimation accuracy of the spectral width $\tau_{c}^{-1}$
of an Lorentzian bath with a quantum probe freely evolving (left)
and dynamically controlled with a CPMG sequence of $N=8$ $\pi$-pulses
(right). The minimal error per measurement $\varepsilon(\tau_{c},t_{opt})$
is much smaller for a qubit probe under optimal dynamical control
than under free evolution (middle).}
\end{figure}

Dynamical control of the qubit probe can drastically improve the estimation
of $\tau_{c}$: as shown in Fig. \ref{fig:2}, sequences of equidistant
$\pi$-pulses (phase-flips), known as Carr-Purcell-Meiboom-Gill (CPMG)
sequences \cite{Carr1954,Meiboom1958,abragam1961principles,Slichter1990},
give rise to a minimal error estimation that obeys $\varepsilon_{CPMG}(\tau_{c},t_{opt})\simeq\frac{2.5}{\sqrt{N_{m}}}$,
while under free evolution the error grows with $g\tau_{c}$, $\varepsilon_{free}(\tau_{c},t_{opt})\propto\frac{g\tau_{c}}{\sqrt{N_{m}}}$,
and is therefore much larger for $g\tau_{c}\gg1$, \emph{i.e.} for
distinctly non-Markovian bath spectra.

\subsection{Bath-mediated transfer of quantum information\label{sub:2.3_Bath-mediated-transfer-of}}

The ability to transfer an unknown quantum state between nodes where
the quantum information (QI) can be reliably stored and/ or processed
is at the heart of QI processing and communication schemes. Since
practically any medium connecting distant nodes corrupts the QI \cite{zwick_robustness_2011,zwick_spin_2012,Zwick_Chapt_2013},
one commonly resorts to probabilistic quantum repeaters \cite{duan2001long},
effected by conditional measurements \cite{Harel96}: only the desired
outcomes are kept while the undesired outcomes are discarded. Such
protocols \cite{duan2001long} are severely constrained by high qubit-overhead
and long average duration of successful QI transfer. It is clearly
desirable to resort to deterministic protocols whenever possible.
Here we advocate the possibility of such protocols, whose high success
rate relies on dynamical control that is optimally adapted to the
medium \cite{Zwick_Optimized_2014,escher_optimized_2011,bensky_optimizing_2012}.

The idea is to write the full Hamiltonian as 

\begin{equation}
H=H_{S}+H_{B}+H_{SB}.
\end{equation}
Here the system $S$ consists of the two spins that constitute the
nodes between which the QI is transferred and a mode (channel) of
the medium that couples these spins, labeled by $k=z$, all other
modes being treated as a thermal bath $B$ to which S is coupled. 

\begin{figure}
\begin{centering}
\includegraphics[width=0.9\textwidth]{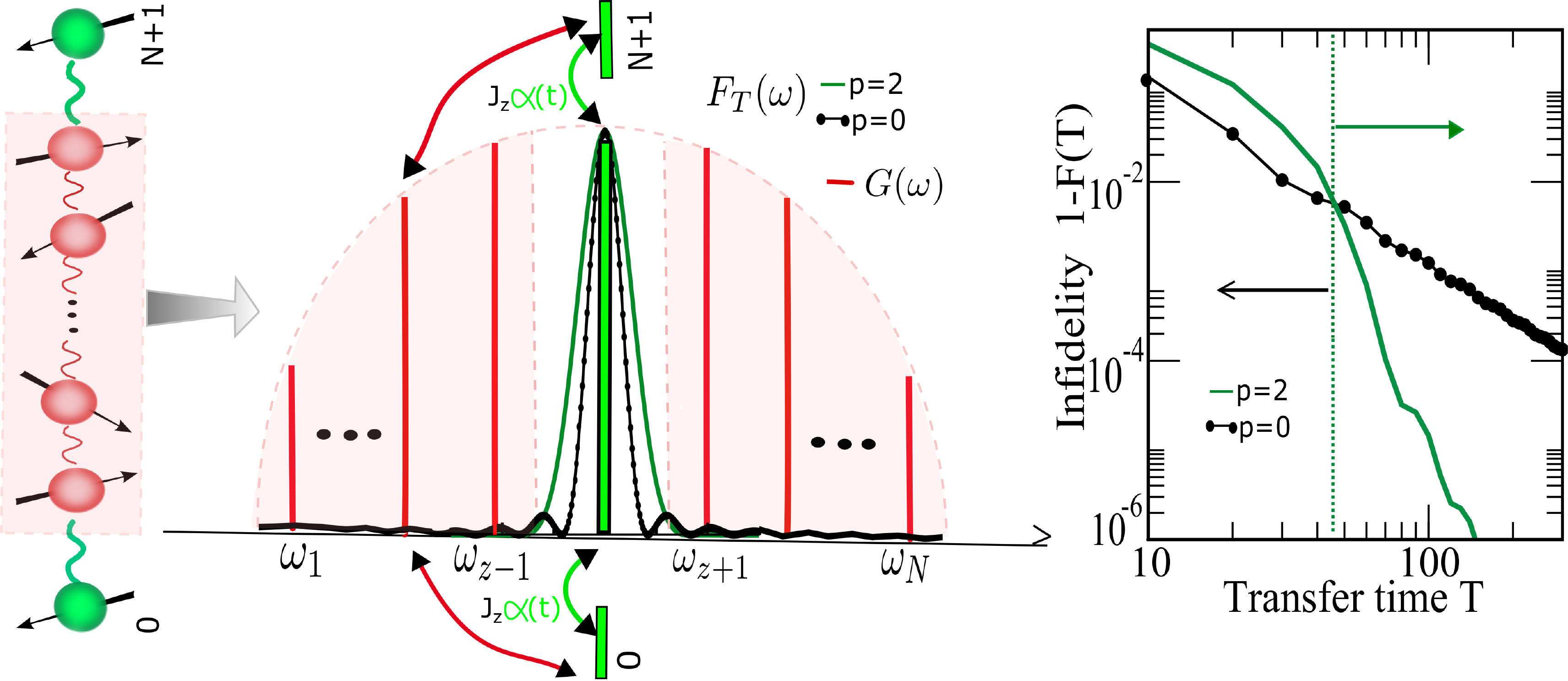}
\par\end{centering}

\protect\caption{\label{fig:3}(Left) Schematic view of quantum information transfer
through a fluctuating quantum spin-channel with random coupling $J$.
The transfer over time $T$ is optimized by an appropriate modulation
$\alpha(t)$, $|\alpha(t)|\le1,$ of the boundary qubit-couplings
to their nearest-neighbors. (Middle) Schematic view of the coupling
spectrum $G(\omega)$ between the ``system'' (comprised of the two
boundary qubits $0$ and $N+1$ and the spin-chain mode that couples
them) and all other modes of the spin chain viewed as a bath (red)
and the filter $F_{T}(\omega)$ corresponding to the optimal chosen
modulation shape $\alpha(t)\propto sin^{2}\left(\frac{\pi t}{T}\right)$
(green) with chopped off tails, as compared to the (free evolution)
filter (black), which has extended tails and therefore much larger
overlap with $G(\omega)$. (Right) The chopping-off of the tails in
the (green) filter may reduce by several orders of magnitude the transfer
infidelity (error) as well as the time required for transfer.}
 
\end{figure}

The transfer fidelity over time $T$ is again governed by the KK universal
formula \cite{kof04,Zwick_Optimized_2014}
\begin{equation}
F(T)\approx1-\int_{-\infty}^{\infty}d\omega F_{T}(\omega)G(\omega).
\end{equation}
We need not know the detailed spectral distribution of the $S-B$
coupling $G(\omega)$, only its width $\mbox{\ensuremath{\frac{1}{\tau_{c}}}}$
and crude mode spacing, which can be estimated by the methods of Secs.
\ref{sub:2.1_Control-for-bath}-\ref{sub:2.2_Maximized-information-on}.
Such estimation should suffice for designing the optimal tradeoff
of the fidelity $F$ versus time transfer $T$ by appropriate temporal
modulation of the coupling. Strikingly, one may analytically prove
\cite{Zwick_Optimized_2014}, upon parameterizing the modulation 
\begin{equation}
\alpha(t)=\alpha_{0}sin^{p}\left(\frac{\pi t}{T}\right)\:p=0,1,2\;|\alpha_{0}|\le1
\end{equation}
that the best tradeoff is usually achievable for $p=2$, because it
yields a filter function without spectral tails (these vanish abruptly
at a controllable frequency). This chopping-off of the tails may reduce
by several orders of magnitude the transfer infidelity (error) as
well as the time required for transfer (Fig. \ref{fig:3})!

This control method, which is universally applicable to media consisting
of interacting fermions (spin-$\frac{1}{2}$ particles) and bosons
alike, may also be used to maximize the storage time of QI inside
a bath memory embodied by an inhomogeneously broadened and thermally
fluctuating spin ensemble \cite{bensky_optimizing_2012}. Thus, the
coupling between quantum systems via the bath is required for effecting
QI transfer or storage, and system-bath interaction control may serve
as a tool for optimizing these processes, on the basis of minimal
knowledge concerning the bath. This method is another application
of our universal procedure for fidelity optimization of the task at
hand and the ability to prioritize the use of resources for implementing
it in any given bath. 

This method may be beneficial for the optimization of operating hybrid
processors of quantum information comprised of different modules \cite{petrosyan_reversible_2009,BenskyQIP11,Kurizki31032015}:
superconducting qubits coupled via a microwave resonator to ensembles
of ultra-cold atoms or NV-center spins. Hybrid processors may profit
from the advantages and make up for the shortcomings of the individual
modules \cite{Kurizki31032015}. Specifically, the superconducting
qubits are fast but vulnerable to decoherence. The outcome of their
operations should be controllably transferred to collective \textquotedblleft quiet\textquotedblright{}
(decoherence-resilient) states of the atoms that are much better suited
for long-term shelving (storage) of this quantum information (QI).
The overall fidelity of the processor can be improved by dynamical
control that optimizes this QI transfer from the noisy to the quiet
(storage) module. Remarkably, for a given energy of the transfer pulse,
the shortest pulse is by no means optimal \cite{escher_optimized_2011}!

A related method can significantly improve the performance of quantum
memories based on spectrally inhomogeneous spin ensembles \cite{bensky_optimizing_2012}.
This method preselects an optimal portion of the ensemble by appropriate
microwave pulse designs.

\section{Bath-induced entanglement in open systems\label{sec: 3_Bath-induced-entanglement}}

Environment effects generally hamper or completely destroy the \textquotedblleft quantumness\textquotedblright{}
of any complex device. Particularly fragile against environment effects
is quantum entanglement (QE) in multipartite systems. This fragility
may disable quantum information processing and other forthcoming quantum
technologies \cite{nie00a,Gisin_Quantum_cryptog_2002,Sergienko_2005_QCC,giovannetti_2011_advances,wolfgramm_2013_entanglement,schmidt_2005_spectroscopy,hempel_2013_entanglement,Kurizki31032015}:
interferometry, metrology and lithography. Commonly, the fragility
of QE rapidly mounts with the number of entangled particles and the
temperature of the environment (thermal \textquotedblleft bath\textquotedblright ).
This QE fragility has been the standard resolution of the Schroedinger-cat
paradox \cite{RevModPhys75715,HarochExploring}: the environment has
been assumed to preclude macrosystem entanglement. But is it inevitable
that Schroedinger cats die of decoherence (as commonly believed \cite{bookBreuer,RevModPhys75715})?
Or, conversely, can a cat be both dead and alive in a thermal bath? 

\begin{figure}
\begin{centering}
\includegraphics[width=0.55\textwidth]{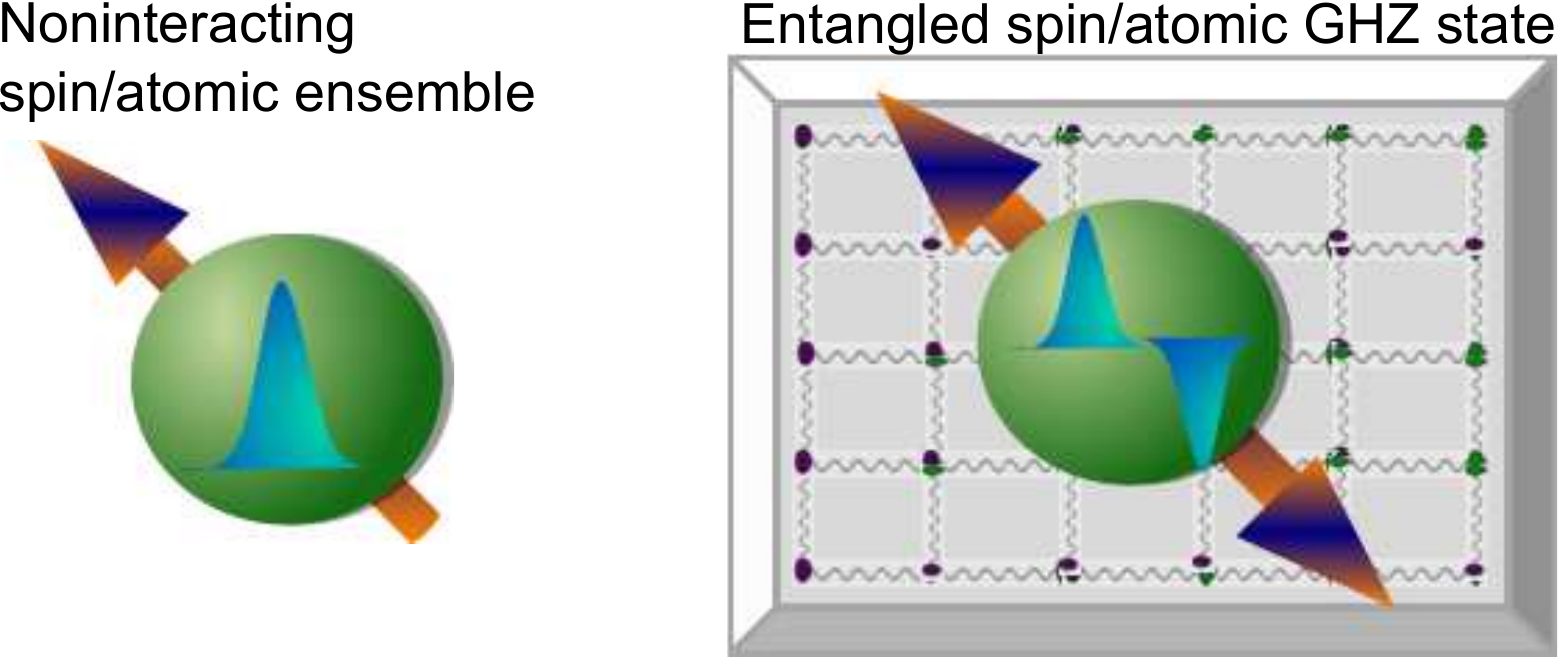}
\par\end{centering}

\protect\caption{\label{fig:4}Schematic view of a product-state spin-polarized ensemble
(left) that spontaneously evolves in the bath into an entangled MQS
or GHZ (Schroedinger-cat) state at a particular time, as a result
of bath-induced entanglement. }
\end{figure}

We shed light on these fundamental issues within the simple model
of N spin-$\frac{1}{2}$ non-interacting particles that identically
couple to a thermal oscillator-bath via the z-component of their Pauli
operators. A single spin in such a model undergoes bath-induced pure
dephasing \cite{bookBreuer,RevModPhys75715,Gonzalo10}. Yet, strikingly
\cite{2011PhRvL.106a0404B,2011PhRvL.107a0404B}, an initial product
state of $N$ $z$-polarized spins can spontaneously evolve via such
coupling to the bath, into a Schroedinger-cat state, also known as
a macroscopic quantum superposition (MQS) or GHZ state \cite{HarochExploring},
nearly deterministically (Fig. \ref{fig:4}) 
\begin{equation}
\left|\Uparrow\right\rangle \longrightarrow p\underset{GHZ}{\underbrace{\left(\frac{\left|\Uparrow\right\rangle +e^{i\frac{\pi}{2}}\left|\Downarrow\right\rangle }{\sqrt{2}}\right)\left(\frac{\left\langle \Uparrow\right|+e^{-i\frac{\pi}{2}}\left\langle \Downarrow\right|}{\sqrt{2}}\right)}}+(1-p)\rho_{S},\:p\simeq1
\end{equation}
with $\left|\Uparrow\right\rangle =\left|\uparrow\uparrow..\uparrow\right\rangle $,
$\left|\Downarrow\right\rangle =\left|\downarrow\downarrow..\downarrow\right\rangle $
and with only a small probability $1-p$ of evolving into an incoherent
state of the $N$ spins.

This dynamics of the collective spin along $z$, $L_{z}=\sum_{j}\sigma_{zj}$,
is driven by the Hamiltonian 
\begin{equation}
\begin{array}{c}
H=\omega_{0}L_{z}+\sum_{k}\omega_{k}b_{k}^{\dagger}b_{k}+\underset{H_{I}=L_{z}B:\:\textrm{collective coupling to bath}}{\underbrace{L_{z}\sum_{k}\eta_{k}(b_{k}+b_{k}^{\dagger})}}\end{array}
\end{equation}
where $\omega_{0}L_{z}$ stands for the collective energy (without
the bath), $b_{k}$ and $b_{k}^{\dagger}$ respectively annihilate
and create bath quanta in modes labeled by $k$, with frequencies
$\omega_{k}$ and their coupling constants to $L_{z}$ are denoted
by $\eta_{k}$. The evolution of the combined system-bath state is\emph{
exactly soluble} by means of the unitary evolution operator \cite{2011PhRvL.106a0404B,2011PhRvL.107a0404B}
\begin{equation}
U(t)=exp\left(-i\left(\omega_{0}tL_{z}+\underset{\textrm{collective Lamb shift}}{\underbrace{f(t)L_{z}^{2}}}\right)+L_{z}t\sum_{k}(\alpha_{k}b_{k}+\alpha_{k}^{*}b_{k}^{\dagger})\right)\label{eq:U(t)}
\end{equation}
Here the bath-induced nonlinear term is a collective Lamb shift (whose
time-dependence is described by the bath-dependent function $f(t)$).
This term does not affect the bath and only entangles the spins. The
last term (wherein $L_{z}$ is coupled to a linear combination of
$b_{k}$ ($b_{k}^{\dagger}$) with amplitudes $\alpha_{k}$) gives
rise to system-bath entanglement which, upon tracing out the bath,
decoheres the spin-system state.

Clearly, the ability to entangle the spins via the bath-induced quadratic
$L_{z}^{2}$ Lamb shift requires the suppression of the decoherence-inducing
term (linear in $L_{z}$) in Eq. (\ref{eq:U(t)}). This suppression
is achievable by control, consisting of periodic phase flips that
tend to average out the linear (odd-symmetry) term but leave intact
the quadratic (even-symmetry) Lamb-shift term.

\begin{figure}
\begin{centering}
\includegraphics[width=0.6\columnwidth]{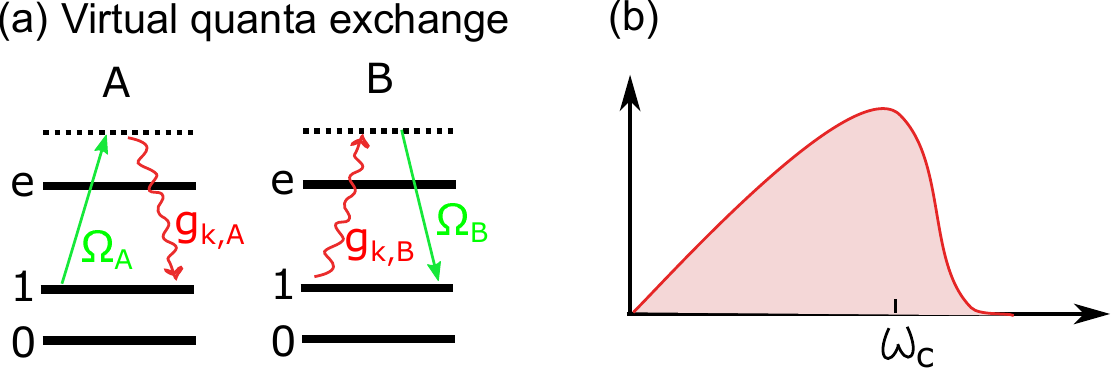}
\par\end{centering}

\protect\caption{\label{fig:5}(a) Schematic view of bath-induced virtual quanta (wiggly
arrows) exchanged between atoms $A$ and $B$, in the presence of
off-resonant fields (solid arrows). The net result is a collective
interatomic energy (Lamb) shift. (b) An Ohmic bath spectrum allows
for a collective Lamb shift associated with the integral over all
bath frequencies $\omega$, from $0$ to $\omega_{c}$, and thus dominates
over the decoherence rate associated with the bath at $\omega\simeq0$.}
\end{figure}

Further insight into the competing effects of bath-induced entanglement
and decoherence can be obtained from a detailed consideration of a
realistic model: two-atom dispersive coupling to a common cavity bath
(Fig. \ref{fig:5}a), described by the interaction Hamiltonian

\begin{equation}
H_{I}=\underset{j=A,B}{\sum}\sigma_{z,j}B_{j}=\underset{j=A,B}{\sum}\underset{k}{\sum}\frac{\Omega_{j}g_{k,j}}{\Delta_{j}}\left|1_{j}\right\rangle \left\langle 1_{j}\right|(b_{k}+b_{k}^{\dagger})
\end{equation}
Here the energy shift of state $\left|1\right\rangle $ in atom \emph{j}
is caused by the combined effect of an off-resonant classical field
(with Rabi frequency $\Omega_{j}$ and detuning $\Delta_{j}$) and
the quantized cavity field (with coupling strength $g_{k,j}$) (Fig.
\ref{fig:5}a). The cross-coupling of atoms $j=A,B$ via virtual quanta
exchange in the cavity is the source of their collective Lamb shift.
This cross-coupling is chosen not to depend on the interatomic distance
under the assumption of identical couplings of both atoms to all cavity
modes:
\begin{equation}
\eta_{k}=\frac{\Omega_{j}g_{k,j}}{\Delta_{j}}
\end{equation}
which is the case for atoms located at symmetric positions in the
cavity. Then the foregoing analysis yields the real-quanta exchange
rate between the atoms that causes decoherence
\begin{equation}
\Gamma_{A(B)}=2\pi G_{T}(\omega=0)
\end{equation}
where the coupling spectrum of the cavity-bath at temperature $T$
is sampled at $\omega=0$. This decoherence rate competes against
the collective Lamb shift 
\begin{equation}
f_{AB}=P\int d\omega\frac{G(\omega)}{\omega}.
\end{equation}
This two-atom Lamb shift is given by the principal-value part of the
integral over the entire coupling spectrum, which, remarkably, is
taken to be at zero temperature, $T=0$, regardless of the actual
bath temperature.

The desired dominance of the collective Lamb shift due to virtual
quanta exchange over decoherence due to real quanta exchange, e.g.,
in an Ohmic bath, holds if

\begin{equation}
f_{AB}\gg\Gamma_{A(B)}\:\textrm{if}\:\omega_{c}\gg G_{T}(0).
\end{equation}
Namely, the upper cutoff frequency far exceeds the zero-frequency
coupling rate, which is typically the case (Fig.\ref{fig:5}b).

We thus arrive at the following paradigms: \emph{(i)} QE in large
multipartite systems may naturally (spontaneously) arise (albeit over
limited time) when the system is embedded in commonly encountered
thermal environments (baths). This QE may yield the spontaneous formation
of Schroedinger-cat states (MQS). \emph{(ii)} QE control may actually
take advantage of the coupling to the environment rather than try
to eliminate it, \emph{i.e.}, it should enhance the \textquotedblleft helpful\textquotedblright{}
coupling, leading to virtual quanta exchange, and suppress the \textquotedblleft harmful\textquotedblright{}
exchange of real quanta via the bath.

Such natural, yet unitary, evolution within thermal baths of the system
to a highly-nonclassical MQS state is a universal effect which we
dub bath-induced entanglement (BIE). Whereas, as a rule, the interaction
of quantum system with a thermal bath gives rise to decoherence, BIE
arises from nonresonant (virtual) interactions between particles via
the bath: nonlinear frequency pulling. This is a generalization of
effects that have previously been studied for multi-ion coupling to
single-mode phonons \cite{Molmer_Sorensen_PhysRevA.62.022311}.

A complementary (orthogonal) approach taken by other groups is to
realize certain entangled states by engineering the incoherent (nonunitary)
dissipation of quantum systems into a Markovian (spectrally flat)
bath \cite{myatt2000decoherence,lin2013dissipative}. By contrast,
the coherent, bath-induced evolution discussed above crucially depends
on having a non- flat bath spectrum. 

Large ensembles of two-level atoms as considered above may be isomorphic
to spin systems with large-spin eigenstates. The interaction of such
ensembles with a common light field may lead to their entanglement
\cite{julsgaard2001experimental}. Here, instead, we rely on the spectra
of commonly encountered baths to drive the spin ensemble into an entangled
state, via effectively nonlinear dynamics. 

At the same time, we must be concerned with the protection of such
bath-induced entangled states from the disentangling effects of other
baths that constitute their environment. Such protection presents
a challenge: how to optimally control multiqubit entangled states?
Our ability to face this challenge relies on our universal approach
to multipartite decoherence control \cite{gor06b,JPB,gor07PRA,gordon_scalability_2011,clausen_bath-optimized_2010,clausen_task-optimized_2012}
(see above).
\begin{figure}
\centering{}\includegraphics[scale=0.5]{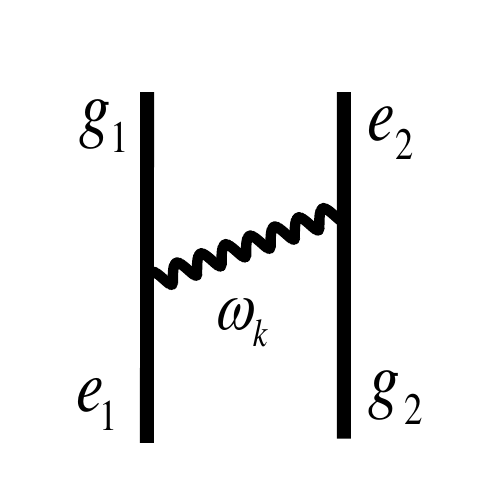} \protect\caption{{\small{}\label{fig:6} Photon-induced interaction between identical
two-level atoms. Atom 1, initially in its upper state $|e_{1}\rangle$,
emits a quantum at mode $k$ and corresponding frequency $\omega_{k}$
while making a transition to its lower state $|g_{1}\rangle$. Atom
2, initially in its lower state, becomes excited upon absorbing the
quantum. When the exchanged photon is real, }\emph{\small{}i.e.}{\small{}
for $\omega_{k}=\omega_{a}$ where $\omega_{a}$ is the frequency
of the atomic transition, the interaction gives rise to dissipative,
and hence probabilistic, cooperative-emission. The summation over
all other virtual-photon-mediated processes, }\emph{\small{}i.e.}{\small{}
over all transition amplitudes for which $\omega_{k}\protect\neq\omega_{a}$,
yields the quantum-mechanically coherent and hence deterministic exchange
process of resonant dipole-dipole interaction (RDDI). }{\small \par}}
\end{figure}

\section{Long-range bath-induced dispersive interactions\label{sec:4_Long-range-bath-induced-dispersi}}

\begin{figure}
\centering{}\includegraphics[scale=0.5]{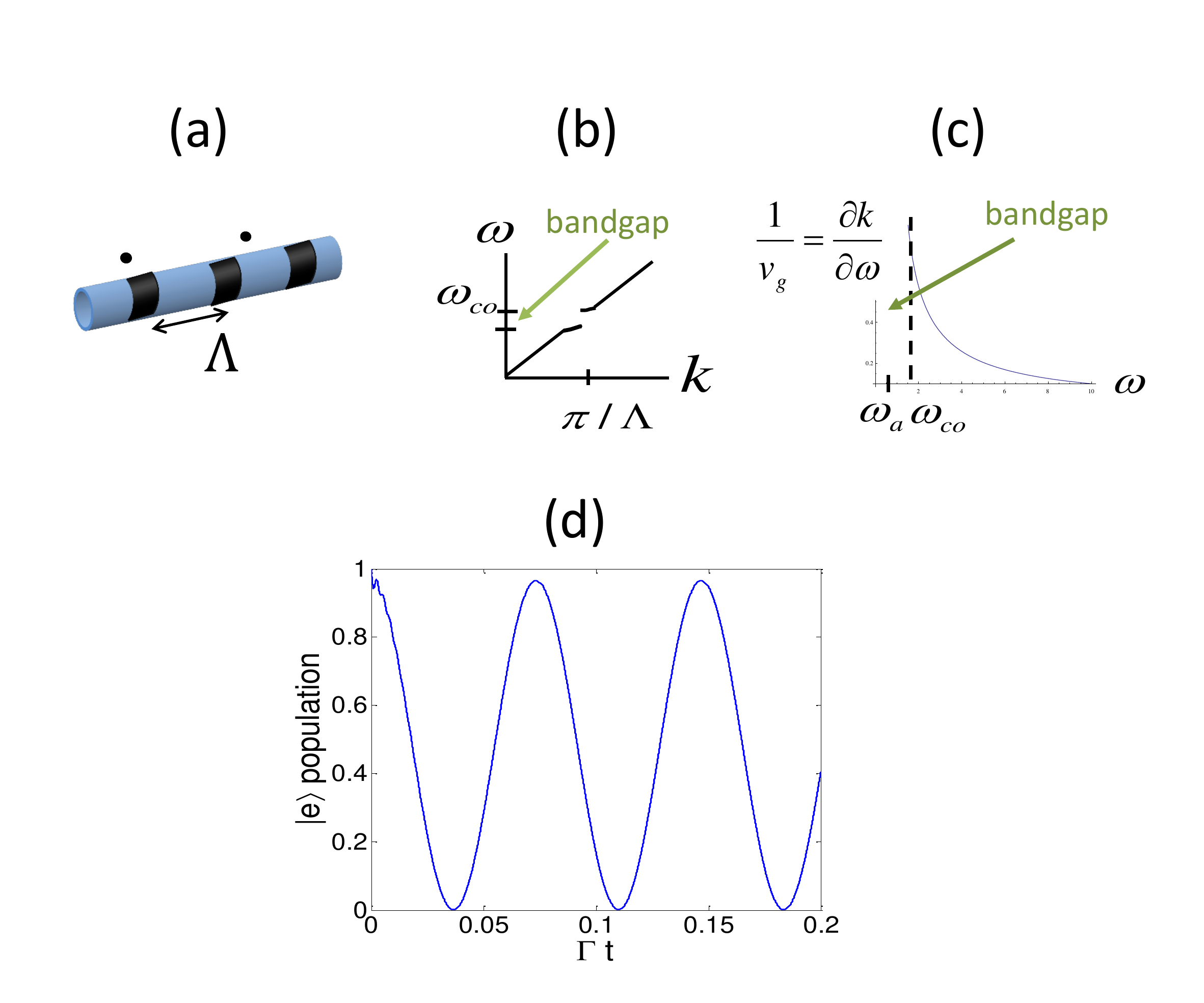} \protect\caption{{\small{}\label{fig:7} Long-range RDDI in a fiber-grating. (a) Atoms
(black dots) coupled to a fiber with a modulated refractive index
(alternating blue and black colors) and a grating period of length
$\Lambda$. (b) Illustration of the dispersion relation of the fiber-grating
$\omega(k)$, $k$ being the photon-mode wavenumber on the fiber axis:
a gap at frequency $\omega$ and $k=\pi/\Lambda$ is opened up, with
the upper bandedge at the upper cutoff frequency $\omega_{co}$. (c)
The density of longitudinal photon modes $\partial k/\partial\omega$
(inverse of group velocity $v_{g}$) vanishes in the gap and diverges
at the bandedge $\omega_{co}$. (d) RDDI mediated by the photon modes
from (c): the excited-state population of atom 1 is plotted using
a non-perturbative theory that goes beyond that of Eq. (\ref{ESRDDI})
\cite{shahmoon2013nonradiative}. This illustration is plotted for
the D1 transition in Rb87 atoms and for an inter-atomic distance of
$z\approx16\mu$m (see \cite{shahmoon2013nonradiative} for more details).
The population, initially unity, slightly decreases to 0.9663 and
then oscillates periodically between 0.9663 and 0, similarly to the
prediction of Eq. (\ref{ESRDDI}). This supports a long-distance entanglement
generation with concurrence $C\approx0.9663$ between the atoms at
a distance of roughly 20 atomic resonant wavelengths, following an
interaction duration of about $t\sim1.8$ns \cite{shahmoon2013nonradiative}.}{\small \par}}
\end{figure}

As argued above, the key to BIE is virtual quanta exchange via the
bath. The BIE processes considered in Sec. \ref{sec: 3_Bath-induced-entanglement}
were restricted to identical coupling of all the atoms to the bath
modes, and hence their collective Lamb shift is distance-independent.
However, in general this is not the case: the system-bath couplings
in the interaction Hamiltonian depend on the positions of the individual
atoms via the spatial mode functions of the bath modes. In free space
the mode functions of the photonic bath are 3d plane waves, giving
rise to real and virtual quanta exchange which both decay with interatomic
separations $r$ and correspond to Dicke-like cooperative emission/absorption
and to cooperative Lamb shifts (\emph{i.e.} resonant dipole-dipole
interaction - RDDI), respectively \cite{Lehmberg_1970_PRA.2.883,Lehmberg_1970_PRA.2.889}.
Whereas for interatomic separations $r$ longer than the resonant
atomic wavelength the real- and virtual-photon processes are comparable
(scaling as $1/r$), in the near-field zone, \emph{i.e.} for small
$r$, the RDDI retains the familiar dipole-dipole scaling as $1/r^{3}$,
and can greatly exceed cooperative decay. Therefore, only in the near-field
zone can free-space RDDI lead to predominantly deterministic BIE.
On the contrary, RDDI-induced entanglement is never deterministic
at separations beyond the emission wavelength, where incoherent absorption
and emission render it probabilistic (Fig. \ref{fig:6}).

\subsection{Long-range deterministic entanglement via RDDI\label{sub:4.1_Long-range-deterministic-entangl}}

\begin{figure}
\centering{}\includegraphics[scale=0.4]{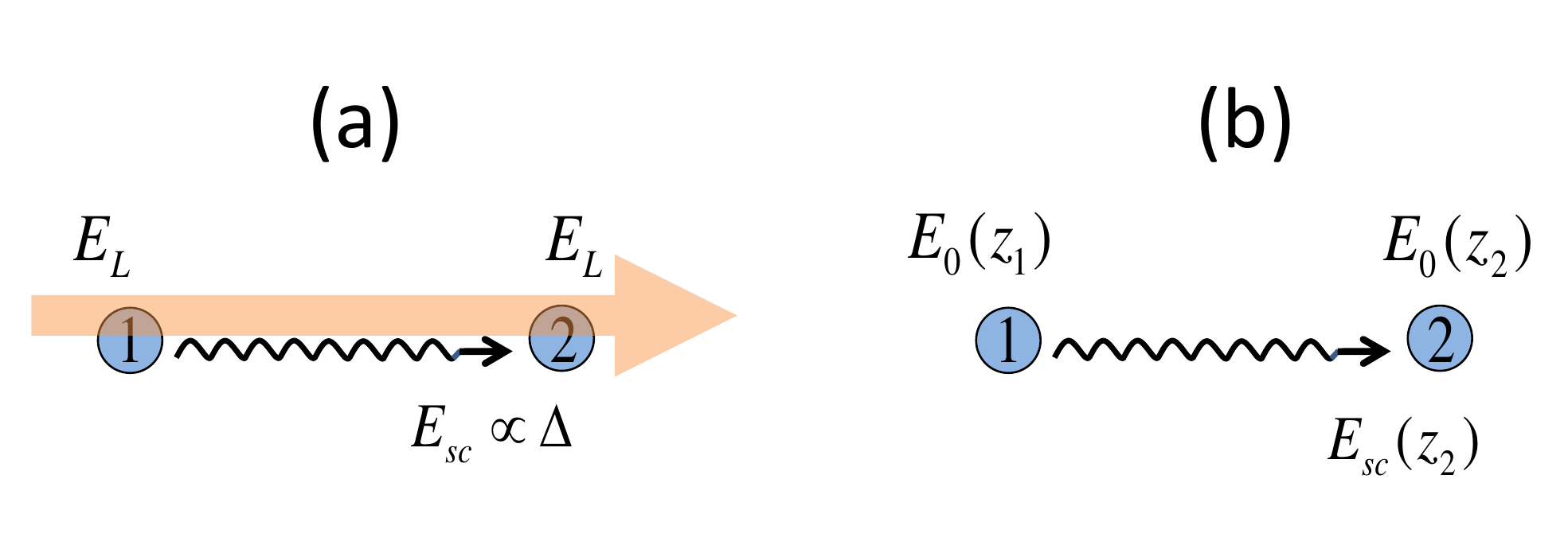} \protect\caption{{\small{}\label{fig:8} Dispersive forces between atoms. (a) Laser-induced
forces: an off-resonant laser field $E_{L}$ illuminates atoms 1 and
2. The scattered field between them, proportional to the RDDI strength
$\Delta$, establishes their interaction. (b) Vacuum-induced forces
(van der Waals and Casimir): here the laser is replaced by the electric
field of the vacuum fluctuations $E_{0}(z)$ (see text).}{\small \par}}
\end{figure}

There is a remedy to this state of affairs that would render BIE via
RDDI nearly deterministic in the far zone, \emph{i.e.} for atoms separated
by many wavelengths. This remedy is based on bath engineering: shaping
photon modes at will by changing the geometry of the bath. The idea
is to consider optical waveguides, such as a Bragg grating, where
the group velocity of guided photonic modes vanishes at the cutoff
(band-edge) frequency, giving rise to giant enhancement of the mode
density \cite{shahmoon2011strongly,shahmoon2013nonradiative,CHA1,shahmoon2013engineering,shahmoon2013dispersion}
(Fig. \ref{fig:7}a,b,c). Then, considering atomic resonance frequencies
within the bandgap but very close to its bandedge (cutoff frequency),
two consequences emerge: 1) The atoms do not exchange real (resonant)
quanta (cooperative decay) along the waveguide due to the vanishing
photon density of states at the atomic resonance, thus eliminating
their probabilistic, dissipative, interaction. 2) The atoms do however,
exchange virtual (nonresonant) quanta via RDDI, mediated by all allowed
waveguide modes (see caption of Fig. \ref{fig:6}). Furthermore, the
resulting RDDI exhibits a strongly enhanced interaction rate (energy)
$\Delta$ and effective range $\xi$, scaling as 
\begin{eqnarray}
 &  & \Delta\propto\frac{\Gamma_{fs}}{\sqrt{1-\omega_{a}/\omega_{co}}},\nonumber \\
 &  & \xi\propto\frac{\lambda_{a}}{\sqrt{1-\omega_{a}/\omega_{co}}}.\label{Efi_D}
\end{eqnarray}

The above expressions reveal a RDDI whose strength $\Delta$ is much
larger than the free-space spontaneous emission rate $\Gamma_{fs}$,
and much longer-range than the atomic resonance wavelength $\lambda_{a}$,
as the atomic resonance frequency $\omega_{a}$ approaches the cutoff
(bandedge) frequency $\omega_{co}$. Since spontaneous emission, being
inhibited into the guided modes, remains at the free-space value whereas
RDDI now has a giant value, we may describe the interatomic exchange
of a photon by the effective Hamiltonian that affects two-atom entanglement
nearly-deterministically, \emph{i.e.}, with high fidelity (Fig. \ref{fig:7}d),
\begin{equation}
H_{eff}=\hbar\Delta{\textstyle \underset{\underset{i\ne j}{i,j=1,2}}{\sum}}\sigma_{i}^{+}\sigma_{j}^{-}\label{ESRDDI}
\end{equation}
where $\sigma_{i}^{+}$ $(\sigma_{j}^{-})$ are the excitation (deexcitation)
Pauli operators of the respective atoms.

\subsection{Long-range laser-induced forces\label{sub:4.2_Long-range laser-forces}}

The same principle that allows for the establishment of coherent BIE
via RDDI while suppressing the incoherent and dissipative process
of emission, can be used to create long-range \emph{conservative forces}
between atoms. When atoms are illuminated by an off-resonant laser,
their motion is affected by two processes in analogy to RDDI and spontaneous
emission: an inter-atomic conservative force, and a diffusive motion
due to the scattering, respectively. The direct relation to RDDI and
spontaneous emission can be seen from the following picture (Fig.
\ref{fig:8}a): the off-resonant laser virtually excites the atoms,
which, once excited, can either interact coherently via RDDI, leading
to a distance-dependent cooperative energy shift and hence a force,
or emit photons, leading to scattering \cite{LIDDI}.

For atoms coupled to a waveguide with a bandgap spectrum as in Fig.
\ref{fig:7}, and illuminated by an off-resonant laser, the resulting
laser-induced force follows the same RDDI strength and range as in
Eq. (\ref{Efi_D}), while the scattering and hence the diffusion,
is suppressed. Therefore, the dynamics of the motion of atoms in such
a system are predominantly affected by an extremely long-range conservative
force. Such a configuration opens the door for the realization and
control of many atoms coupled by long-range forces, that are expected
to exhibit unique thermodynamic features, such as inequivalence of
statistical ensembles and anomalously slow relaxation to equilibrium
\cite{shahmoon2014non}.

\subsection{Long-range vacuum-induced forces\label{sub:4.3_Long-range vacuum-forces}}

\begin{figure}
\centering{}\includegraphics[scale=0.3]{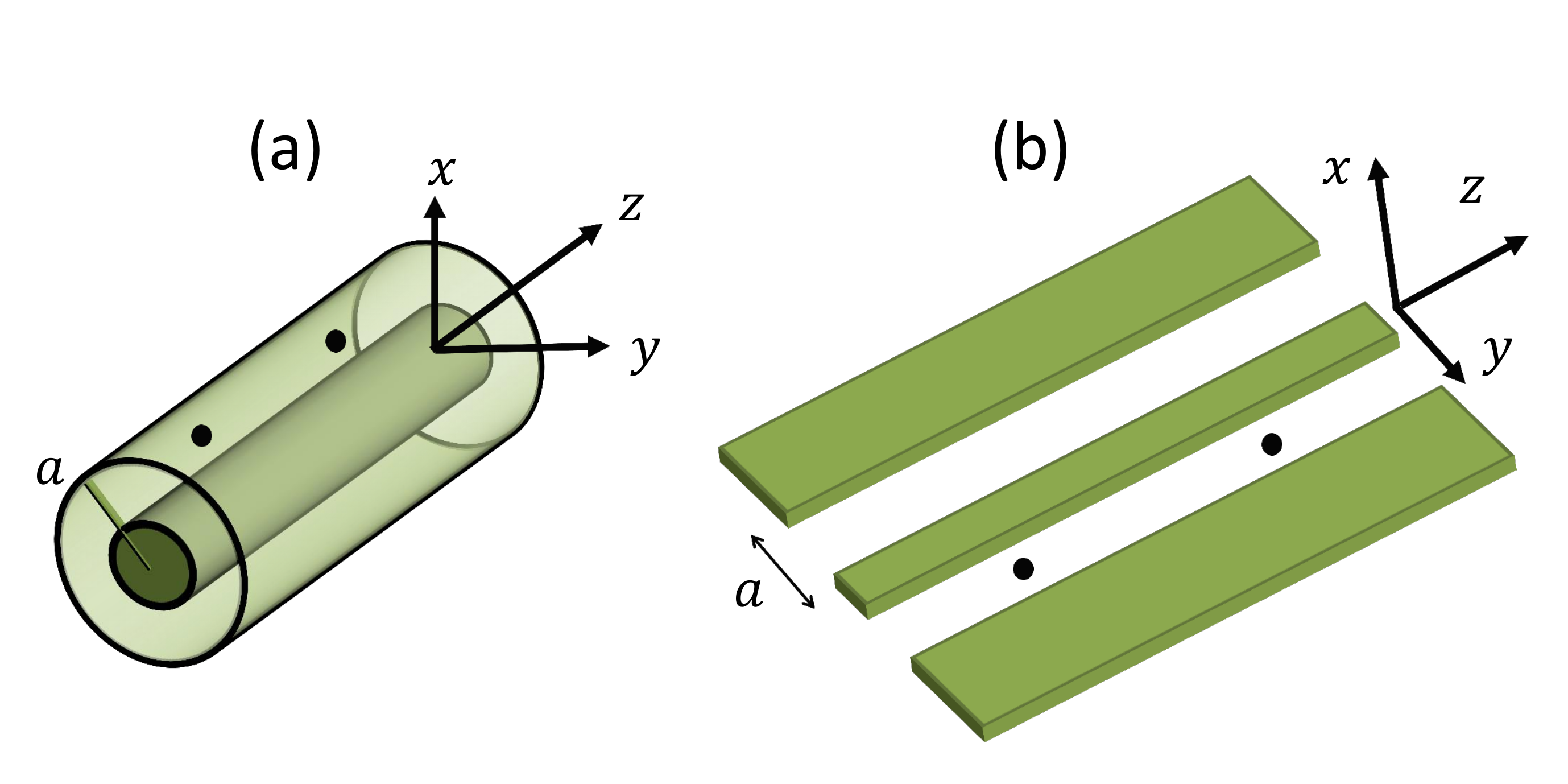} \protect\caption{{\small{}\label{fig:9}Atoms (black dots) coupled to a transmission-line
(TL): a TL is typically comprised of two distinct conductors. (a)
A coaxial TL comprised of two concentric conducting cylinders. (b)
A coplanar waveguide may be seen as an open version of the coaxial
line. It is comprised of a central conductor and two other grounded
conductors (at a distance $a$ from the center).}{\small \par}}
\end{figure}

Even more dramatic, giant, enhancement is achievable via the control
of the bath-geometry, for dipolar forces induced by the electromagnetic
vacuum, namely, the Casimir and van der Waals (vdW) forces. The idea
is to consider atoms coupled to an electric transmission line (TL),
such as a coaxial cable or coplanar waveguide \cite{shahmoon2014giant}
(Fig. \ref{fig:9}), which support the propagation of quasi-1d transverse
electromagnetic (TEM) modes. Then, virtual excitations (photons) of
these extended modes can mediate much stronger and longer-range Casimir
and vdW forces than in free-space.

The unique feature of the fundamental TEM modes is their dispersion-free
and diffraction-free 1d propagation, revealed by the $k-$dependence
of their frequency $\omega_{k}$ and spatial mode function $u_{k}(\mathbf{r})$,

\begin{equation}
\omega_{k}=|k|c,\quad u_{k}(\mathbf{r})=\frac{e^{ikz}}{\sqrt{A(x,y)L}},\label{Efi_u1d}
\end{equation}
$k$ being the wavenumber in the longitudinal waveguide direction
$z$ and $A$ the mode area.

\begin{figure}
\centering{}\includegraphics[scale=0.27]{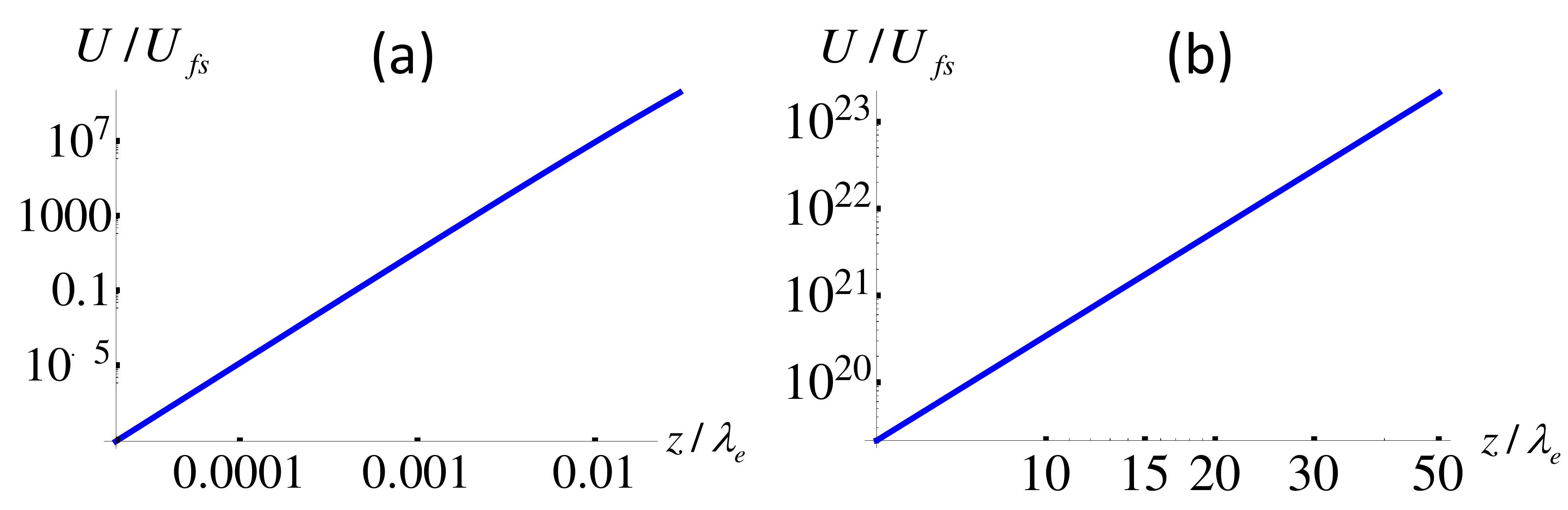} \protect\caption{{\small{}\label{fig:10} TEM-mediated vacuum energy $U$ between a
pair of atoms at a distance $z$, coupled to a TL with transverse
dimension $a=10^{-4}\lambda_{e}$, where $\lambda_{e}$ is a characteristic
dipole-transition wavelength (typical of coplanar-waveguide circuit-QED
setups), compared to its free-space counterpart $U_{fs}$: (a) near-zone
limit $z\ll\lambda_{e}$ (vdW regime); (b) far-zone limit $z\gg\lambda_{e}$
(Casimir-Polder regime). }{\small \par}}
\end{figure}

The contribution of the TEM modes to the Casimir and vdW potentials
can be evaluated by fourth order perturbation theory that yields the
energy shift of two atoms in their lowest (ground) states coupled
to the vacuum of the modes in Eq. (\ref{Efi_u1d}) \cite{MQED}. An
alternative approach consist in recalling the expression for the energy
of an electric dipole, $U=-(1/2)\alpha E^{2}$, where $\alpha$ is
the dipolar polarizability and $E$ is the field at the position of
the dipole \cite{MIL,FoQV}. Considering the energy of, e.g., atom
2, the electric field at its position $z_{2}$ along the transmission
line, is comprised of the ordinary vacuum fluctuations, $E_{0}(z_{2})$,
and those scattered by atom 1 and subsequently arriving at atom 2,
$E_{sc}(z_{2})$. To lowest order in the scattering, this scattered
field is found by the 1d propagation equation of the TEM modes, driven
by the polarization at the location of atom $1$, $E_{0}(z_{1})$
(Fig. \ref{fig:8}b)

\begin{equation}
(\partial_{z}^{2}+k^{2})E_{sc,k}(z)=-\mu_{0}\omega_{k}^{2}\alpha_{1}(\omega_{k})E_{0,k}(z_{1})\delta(z-z_{1})/A,\label{Efi_helm}
\end{equation}
where $k$ is the wavenumber of the field fluctuations. The solution
of the above equation yields $E_{sc,k}(z_{2})$ in terms of $\alpha_{1}(\omega_{k})E_{0,k}(z_{1})$.
On the other hand, the interaction energy between the atoms 1 and
2, deduced from the dipolar energy $U$ of atom 2, is related to the
scattered-field and hence, to lowest order in the scattering, is given
by, 
\begin{equation}
U_{12}\propto-\sum_{k}\alpha_{2}(\omega_{k})\left[E_{sc,k}(z_{2})E_{0,k}(z_{2})+\mathrm{h.c.}\right]\propto-\sum_{k}\alpha_{2}(\omega_{k})\alpha_{1}(\omega_{k})\left[E_{0,k}(z_{1})E_{0,k}(z_{2})+\mathrm{h.c.}\right].\label{Efi_U}
\end{equation}
Finally, treating the vacuum fluctuations $E_{0,k}(z)$ as a quantum
field operator, we average Eq. (\ref{Efi_U}) with respect to the
vacuum state and obtain the vacuum interaction energy between the
atoms, mediated by the dominant TEM mode of the transmission line.

This enhanced interaction energy may be analytically evaluated in
the form of a hypergeometric function $F$ of the interatomic distance
$z=|z_{1}-z_{2}|$ scaled to a typical dipolar-transition wavelength
$\lambda_{e}$. In the near zone, \emph{i.e.} for $z$ much shorter
than this wavelength, we obtain a modified vdW interaction as compared
to free space \cite{shahmoon2014giant} 
\begin{equation}
F(z)\approx\pi+16\pi\frac{z}{\lambda_{e}}\ln\left(\frac{z}{\lambda_{e}}\right),
\end{equation}
that falls off very differently with $z$ than the inverse 6th power,
$U\propto1/z^{6}$, that characterizes the interaction in free space.
By contrast, for far-zone distances (well beyond $\lambda_{e}$),
we find \cite{shahmoon2014giant} 
\begin{equation}
F(z)\approx\frac{1}{(2\pi)^{3}}\left(\frac{\lambda_{e}}{z}\right)^{3},
\end{equation}
as compared to the inverse 7th power falloff in free space, $U\propto1/z^{7}$.
The resulting enhancement can be enormous, as seen in Fig. \ref{fig:10}.

\begin{figure}
\centering{}\includegraphics[scale=0.45]{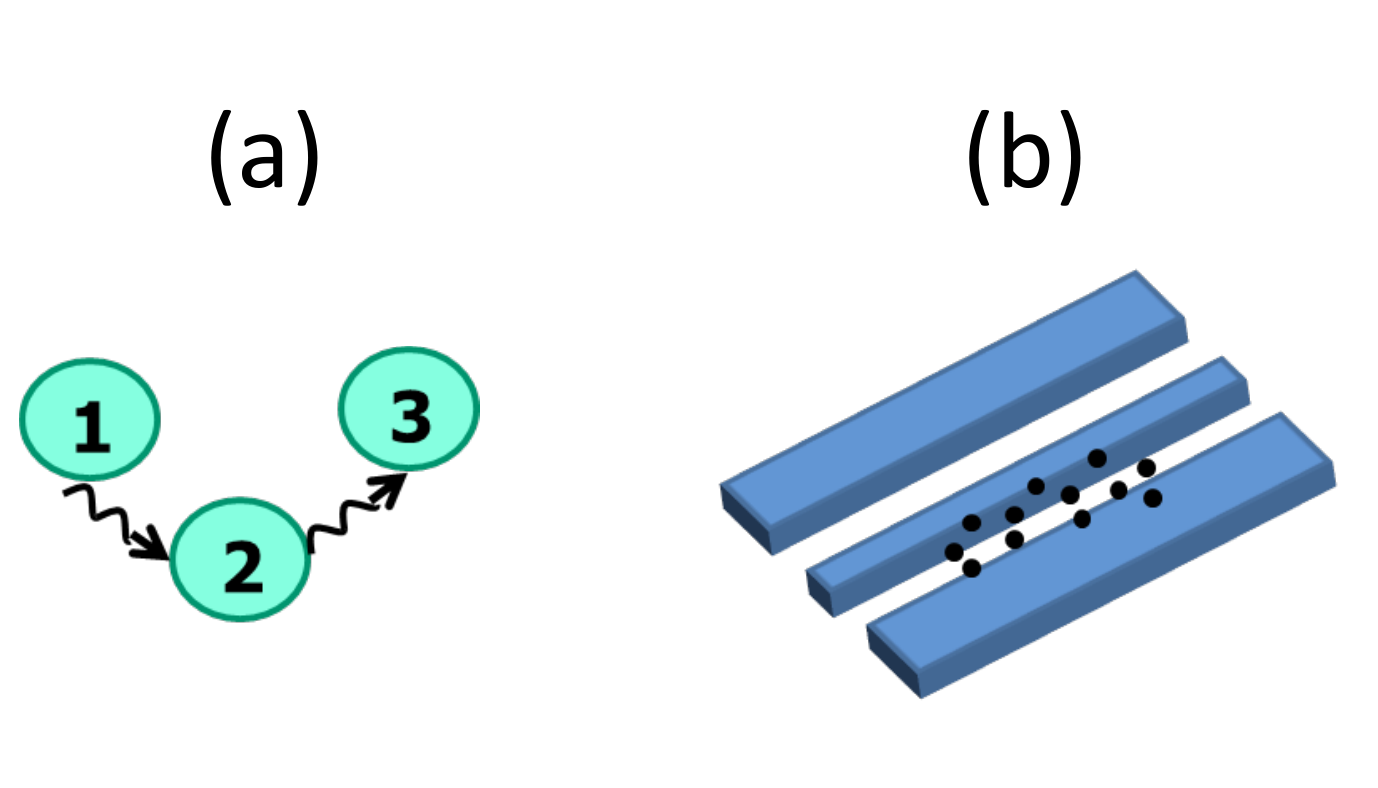} \protect\caption{{\small{}\label{fig:11} Nonadditivity of dispersive interactions.
(a) The total energy of 3 atoms is given by the sum of their pairwise
interactions $\sum_{ij=1}^{3}U_{ij}$ added to their 3-body interaction
$U_{123}$, which is obtained by considering multiple scattering events
involving all atoms. When $U_{123}$ is negligible with respect to
the pairwise summation part, the total energy is called }\emph{\small{}additive}{\small{}.
(b) A gas of atoms coupled to a transmission line, where nonadditive
effects may be observed.}{\small \par}}
\end{figure}

An important outcome of the enhancement of Casimir forces in such
a TL structure may be the onset of \emph{nonadditivity} of atom-atom
interactions \cite{MIL,FoQV,REV} (Fig. \ref{fig:11}a): already at
rather low gas densities we expect that 3-body interactions $U_{123}$
may become comparable to the strength of their usual pairwise counterpart
$U_{12}$, because of the extension of the mediating photon modes
in a TL over long distances. This means that the vacuum energy of
a many-atom system may not be represented by the sum of its pairwise
interactions $U_{12}$. More specifically, the ratio of 3-body to
pairwise interaction energies in free space (3d) scales as \cite{MIL,FoQV}
\begin{equation}
\left.\frac{U_{123}}{U_{12}}\right|_{3d}\propto\frac{\alpha}{z^{3}}.
\end{equation}
Therefore, the inverse of the polarizability $1/\alpha$ sets the
scale for a typical density $\sim1/z^{3}$ where this ratio is large
and nonadditivity is important. For the 1d TEM-mediated case, as in
a TL, we find in the far-zone regime 
\begin{equation}
\left.\frac{U_{123}}{U_{12}}\right|_{1d}\propto\frac{\alpha}{a^{2}z},
\end{equation}
where $a\sim\sqrt{A}$ is transverse dimension of the TL. Then, for
the typical case of $z\gg a$, where the TEM modes are dominant and
1d behavior prevails, nonaddivity is expected to become important
at densities much lower than in free space. Among the possible consequences
of this nonadditivity are drastic modifications of the effective dielectric
response and the heat capacity of gases coupled to such structures
(Fig. \ref{fig:11}b).

\section{Thermodynamic control via quantum Zeno \& anti-Zeno effects\label{sec:5_Thermodynamic-control-via}}

It is clearly desirable to cool down or purify a qubit at the fastest
rate possible to make it suitable for tasks of quantum information
processing. The standard, straightforward way of cooling a system
such as a qubit is by equilibrating this system with a cold bath.
But can one cool qubits faster than their equilibration time?

Another issue concerning cooling is that it often involves transitions
between the qubit levels and other auxiliary levels. But what if such
auxiliary levels are not available? We have shown that these two obstacles
may be overcome by exerting highly frequent perturbations on a qubit,
such as phase shift, or measurements \cite{ere08Nature} at intervals
much shorter than the memory (correlation) time of the bath to which
the qubit couples, and well within the equilibration times \cite{gor10NJP,gor09}.
In such a scenario, a non-Markovian treatment of the purification
must be adopted \cite{ere08Nature,gor10NJP,gor09,Gonzalo10}.
\begin{figure}
\centering{}\includegraphics[width=0.8\textwidth]{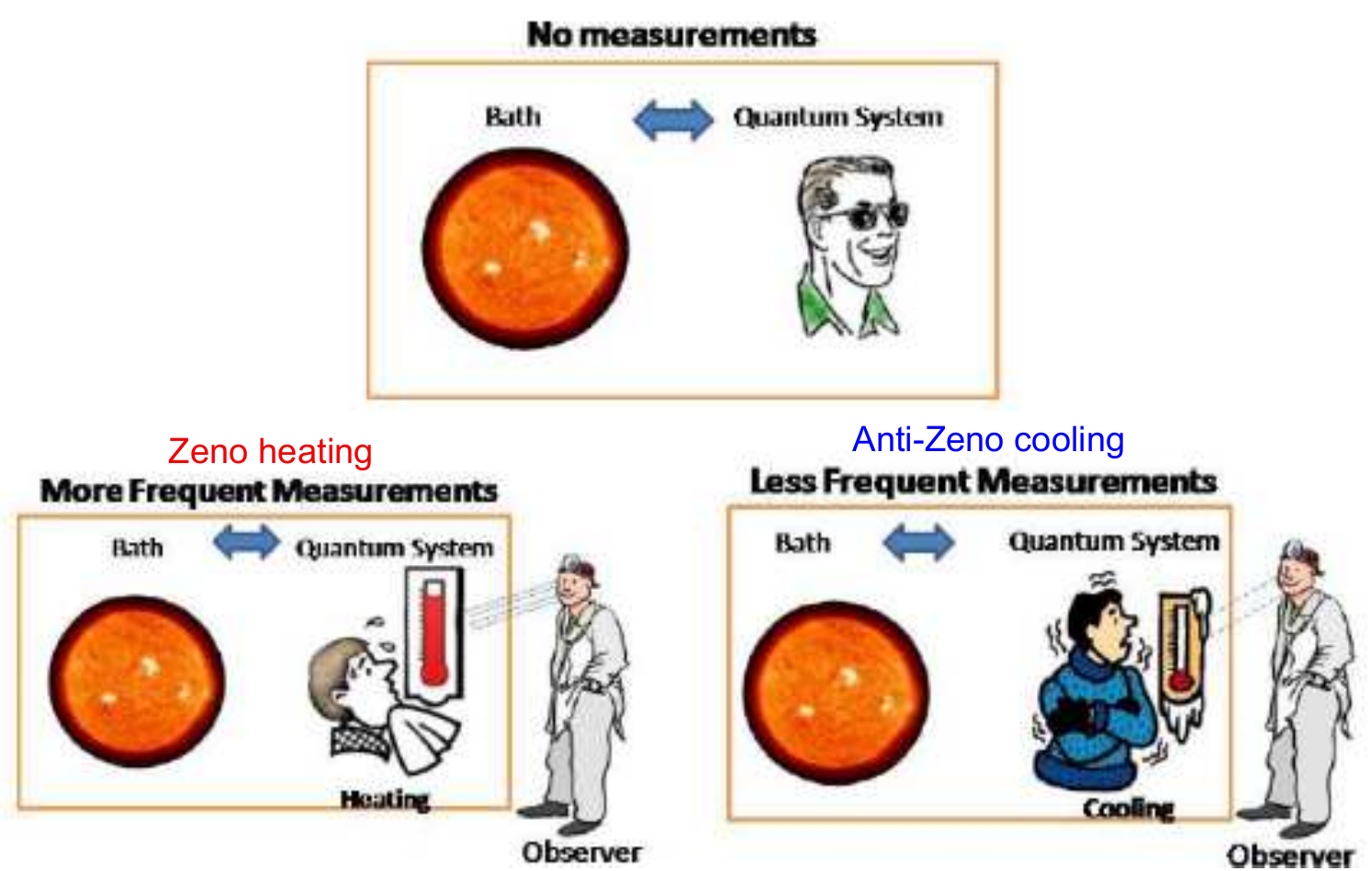}\protect\caption{\label{fig:12}Cartoon of a quantum system whose thermal equilibrium
with a bath (top) is changed towards progressive heating by highly
frequent quantum non demolition (QND) measurements of its energy in
the Zeno regime (bottom left) or towards progressive cooling by less
frequent QND measurements in the anti-Zeno regime (bottom right) \cite{ere08Nature}.}
\end{figure}

\begin{figure}
\centering{}\includegraphics[width=0.65\textwidth]{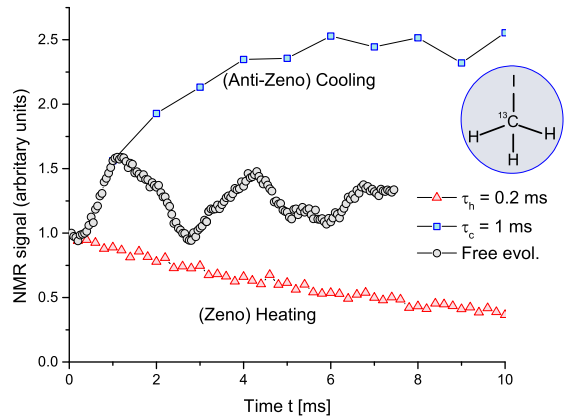}\protect\caption{\label{fig:13}Experimental verification \cite{Gonzalo10} of Zeno
heating for QND polarization measurements (at intervals $\tau=0.2$
ms) and anti-Zeno cooling (at intervals $\tau=1\,m\mbox{s}$), as
in the cartoon above (Fig. \ref{fig:12}) . Experimental setup: quantum
system embodied by the nuclear spin of a carbon atom C is in contact
with a bath embodied by the nuclear spins of 3 protons. Their off-resonant
frequency mismatch is changed by the measurements, causing polarization
decrease (heating) or increase (cooling) of C depending on the interval
$\tau$.}
\end{figure}

Specifically, we have experimentally and theoretically demonstrated
(in collaboration with L. Frydman\textquoteright s group) the possibility
of purifying a qubit coupled to a spin bath, by means of repeated
noise-induced dephasing that mimics the effect of a non-selective
(unread) measurements. We have shown (Fig. \ref{fig:13}) that the
qubit may be cooled down to a predetermined temperature that may be
much lower than that of the bath by means of a suitable controlled
dephasing rate, that conforms to the condition for the anti-Zeno effect
(AZE). By contrast, a dephasing rate that corresponds to a quantum
Zeno effect (QZE) leads to heating of the qubit \cite{ere08Nature,gor10NJP,gor09,Gonzalo10}.
The qubit may exist in the state of predetermined purity to which
it was driven by the measurements or dephasing as long as the entropy
of the bath remains constant. A violation of this (Bonn) approximation
will render the qubit as well as the bath fully mixed, \emph{i.e.}
will thermalize them to infinite temperature \cite{gor10NJP}.

\section{Heat-machine design by system-bath control: quantum thermodynamic
bounds\label{sec:6_Heat-machine-design-by}}

\subsection{Work-information tradeoff in the non-Markovian domain\label{sub:6.1_Work-information-tradeoff-in}}

Open-system manipulations must be optimized within thermodynamic bounds,
concerning entropy, work and heat production. However, when these
manipulations are faster than the bath memory time, so that the Markovian
approximation does not hold, we must revisit and better understand
these bounds. Part of the reason is that correlations between the
system and the bath, which are ignored in standard thermodynamics,
may play an important role on non-Markovian time scales. In particular,
as discussed below, they may invalidate the bound set by Szilard,
and more quantitatively by Landauer, on the tradeoff between information
and work \cite{landauer1961IBM,Szilard1929}.

\begin{figure}
\centering{}\includegraphics[width=0.6\textwidth]{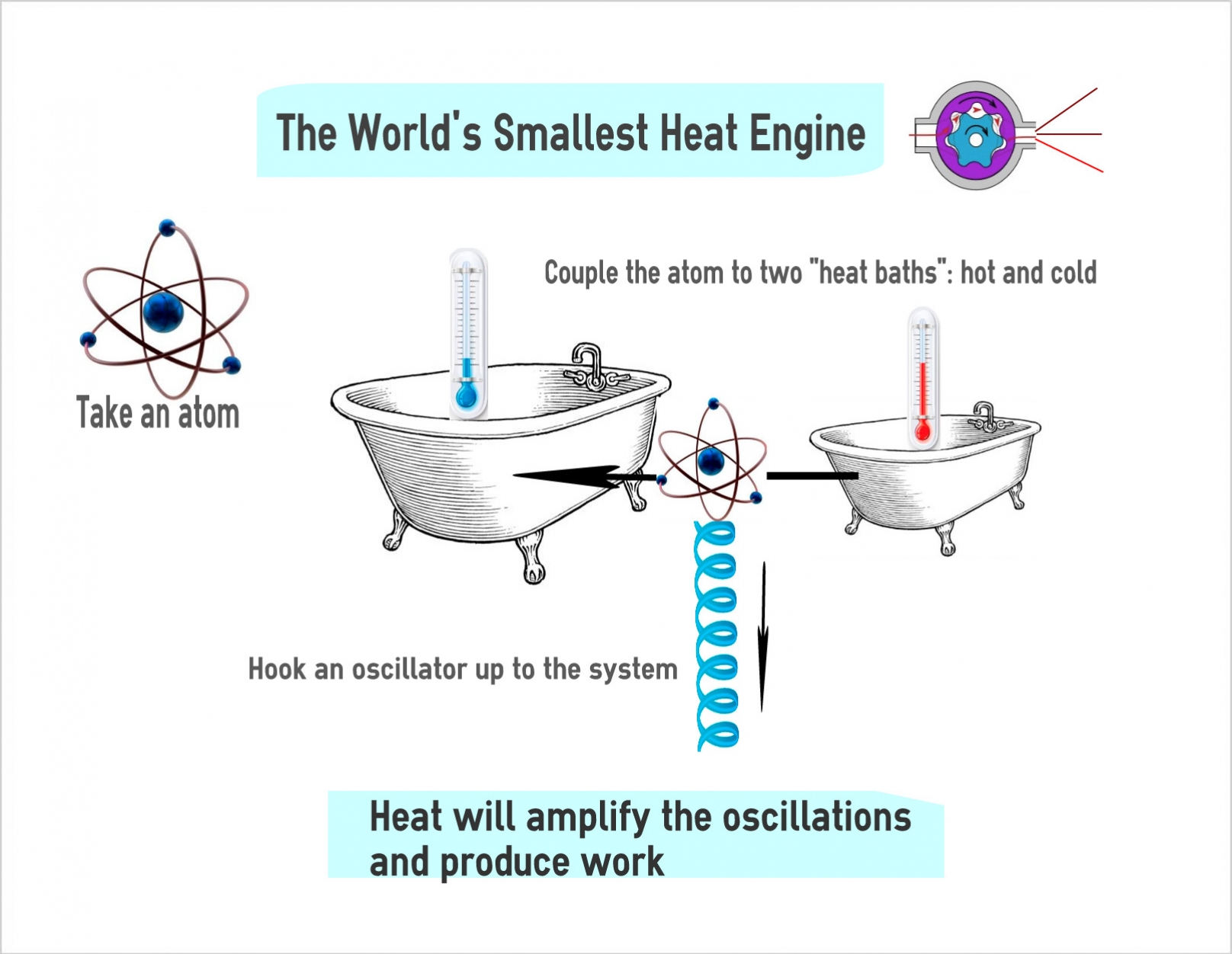}\includegraphics[width=0.6\textwidth]{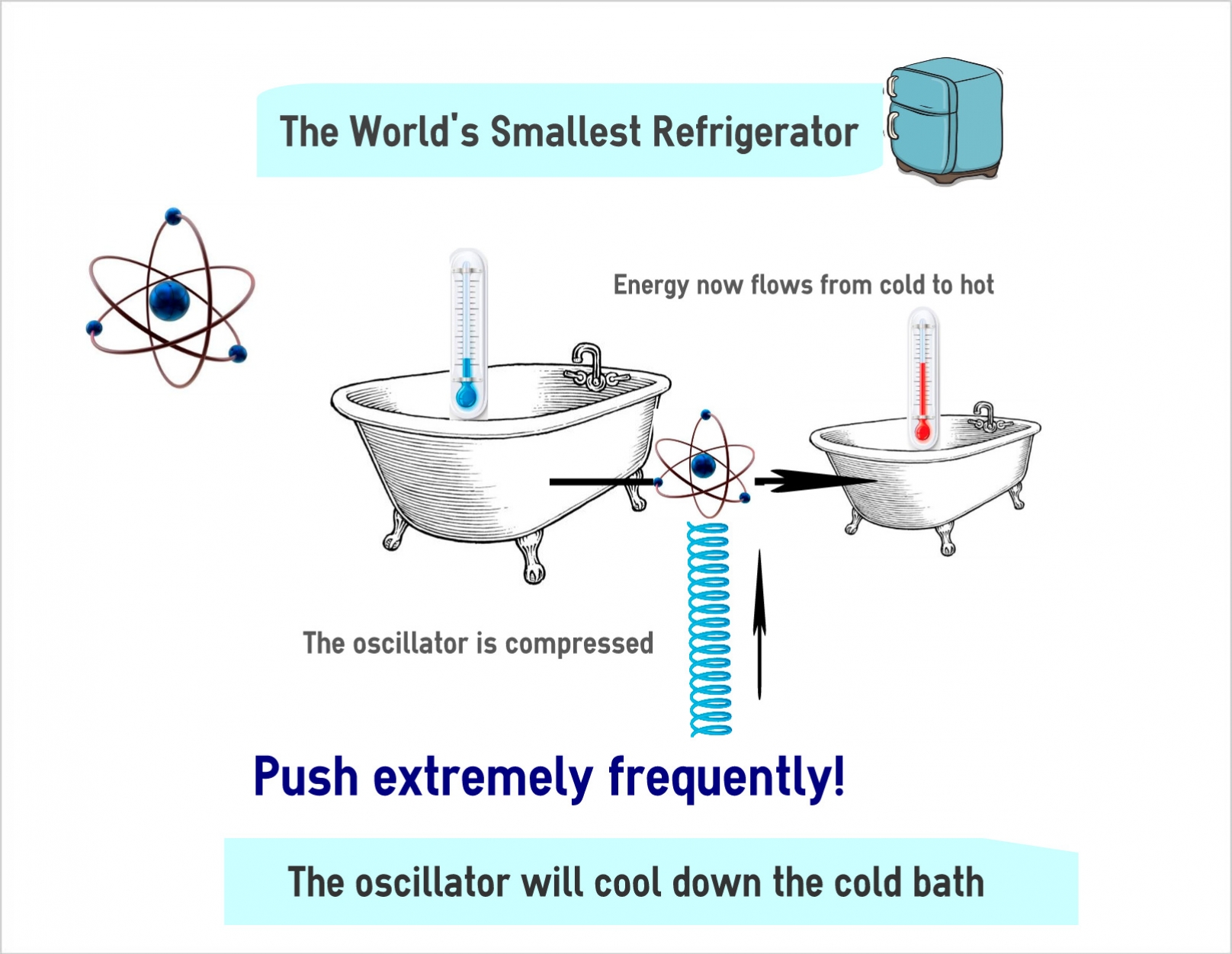}\protect\caption{\label{fig:14}Cartoon of the ``world's simplest and smallest''
universal heat machine acting as heat engine (left) or refrigerator
(right). It is comprised of a qubit that constantly interacts with
two baths at different temperatures. Concurrently, it is driven by
an oscillator that off resonantly modulates (by periodic Stark shifts)
the qubit resonance frequency. The modulation rate determines whether
the qubit will act as a heat engines or refrigerator \cite{gelbwaser2013minimal}.}
\end{figure}

\textbf{System-bath correlation effects.} All existing treatments
of heat engines are based on the assumption that the working-medium
(system) is \emph{autonomous}: its evolution (described by a Lindblad-type
master equation for the bath-averaged system-state $\rho_{S}(t)$)
suffices for a thermodynamic analysis of an engine \cite{alicki1979quantum,Lindbladbook}
driven by Hamiltonian $H_{S}(t)$. Under this standard assumption,
the following expression for the work, $W$, is expected to hold \cite{alicki1979quantum}
for a closed cycle: 
\begin{equation}
W=\oint tr\{\rho_{S}\dot{H}_{S}\}dt.
\end{equation}
The convention is that the work $W$ is negative if it is performed
by the system on the external piston. According to Lindblad's H-theorem,
or the Kelvin formulation of the second law, $W\geq0$, \emph{i.e.}
the system cannot do work on the piston in a single-bath setup. Yet,
strikingly, according to our results \cite{2013PhRvA..88b2112G} if
the cycle is faster than the bath memory time, the system may do net
work on the piston! To resolve the paradox, we contend that, contrary
to the standard assumption \cite{alicki1979quantum,Lindbladbook},
\emph{it is wrong to assume that the system is autonomous in the quantum
non-Markovian domain}: the correlations of the system with the bath
are then crucial! Accordingly, we show that the second law is upheld
if we allow for the energetic and entropic cost incurred upon \emph{decorrelating
the entangled system-bath state} that exists in thermal equilibrium
by the measurement that triggers each cycle. The correct description
must account for the \emph{total} work during the cycle, evaluated
by considering the total state $\rho_{tot}$ and the Hamiltonian $H_{tot}$
(of the system and the bath combined) \cite{2013PhRvA..88b2112G}:
\begin{equation}
W_{tot}=\oint tr\{\rho_{tot}\dot{H}_{tot}\}dt.\label{eq:W_tot}
\end{equation}
\begin{figure}
\centering{}\includegraphics[width=0.55\textwidth]{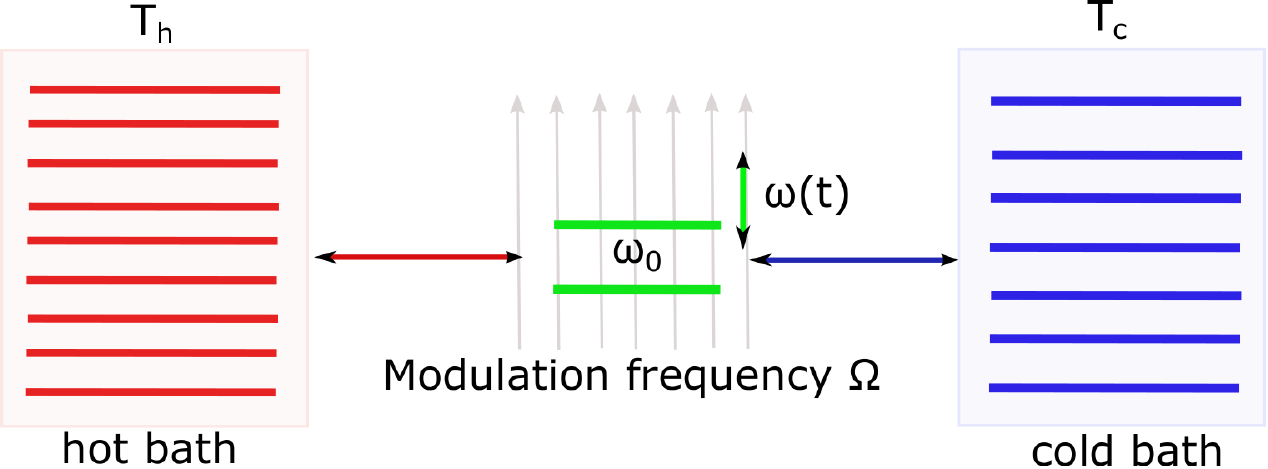}\protect\caption{\label{fig:15}Components of the universal heat machine depicted in
Fig. \ref{fig:14}: (1) working fluid (qubit system); (2) two baths
at different (hot and cold) temperatures, permanently coupled to the
qubit (via weak coupling); (3) a piston (external field oscillator)
that periodically modulates the qubit (level distance), \emph{i.e.}
$\omega(t)$ about the qubit frequency $\omega_{0}$ and with modulation
frequency (rate) $\Omega$, and thereby extracts work (at expense
of the hot bath) in the heat-engine regime, or provides work (in order
to cool down the cold bath) in the refrigerator regime, depending
on $\Omega$ \cite{gelbwaser2013minimal}.}
\end{figure}

The non-negativity of the work $W_{tot}\geq0$, under a closed-cycle
(\textit{unitary}) evolution of the Hamiltonian ($H_{tot}$), in keeping
with the second law, is ensured by the inequality 
\begin{equation}
W_{tot}=W+\Delta E>0,
\end{equation}
where $\Delta E$ is the \emph{measurement} \emph{cost} of the system-bath
decorrelation. Equation~ (\ref{eq:W_tot}) may still allow the system
to do net work during the cycle ($W<0$), but it should compensate
for this work by the energy cost $\Delta E$ of the system-bath state
preparation. This cost comes about from changing the mean system-bath
correlation energy from its negative equilibrium value \cite{ere08Nature,gor10NJP,gor09}
to zero (or positive value) after the preparation (e.g., via a brief
measurement or phase-flip).

\subsection{Minimal quantum heat machines\label{sub:6.2_Minimal-quantum-heat}}

One of our main targets is the strive to realize the minimal and simplest
thermal machines in the quantum domain. These may be conceived as
(Fig. \ref{fig:14}) quantized (harmonic-oscillator) ``piston''
that couples to a single qubit acting as either a quantum heat engine
(QHE) or quantum refrigerator (QR) on spectrally non-flat baths. Floquet
analysis of periodically-driven open systems is used to treat their
steady-state thermodynamics \cite{gelbwaser2013minimal,2012arXiv1205.4552A}.
This formalism aims to separate those distinctly non-Markovian (and
non- rotating-wave) effects that may cause anomalous thermodynamic
phenomena on short-time scales \cite{ere08Nature,gor10NJP,gor09,Gonzalo10}
from steady-state thermodynamics.

The simplest variant to be used as a model for the minimal quantum
thermal machine is a (frequency-modulated) qubit Hamiltonian 
\begin{equation}
H(t)=\frac{1}{2}\omega(t)\sigma_{z},~\omega(t+\frac{2\pi}{\Omega})=\omega(t),
\end{equation}
where $\omega(t)$ is the periodic modulation about the qubit resonance
$\omega_{0}$ with frequency $\Omega$. Its coupling to two heat baths
is given by the interaction Hamiltonian 
\begin{equation}
H_{int}=\sigma_{x}\otimes(B^{h}+B^{c}).
\end{equation}
where the bath operators ($B^{h}$, $B^{c}$) are respectively associated
with hot ($T_{h}$) and cold ($T_{c}$) temperatures, as well as with
\emph{distinct spectra}. Using the period-averaged (coarse-grained)
Floquet expansion, we can find the conditions for the coarse-grained
density operator $\tilde{\rho_{S}}$ of the qubit to be a steady state
of a Lindblad super-operator, 
\begin{equation}
\mathcal{L}\tilde{\rho_{S}}=0,\:\mathcal{L}=\sum_{q}\mathcal{L}_{q}^{a},
\end{equation}
\begin{figure}
\centering{}\includegraphics[width=0.5\columnwidth]{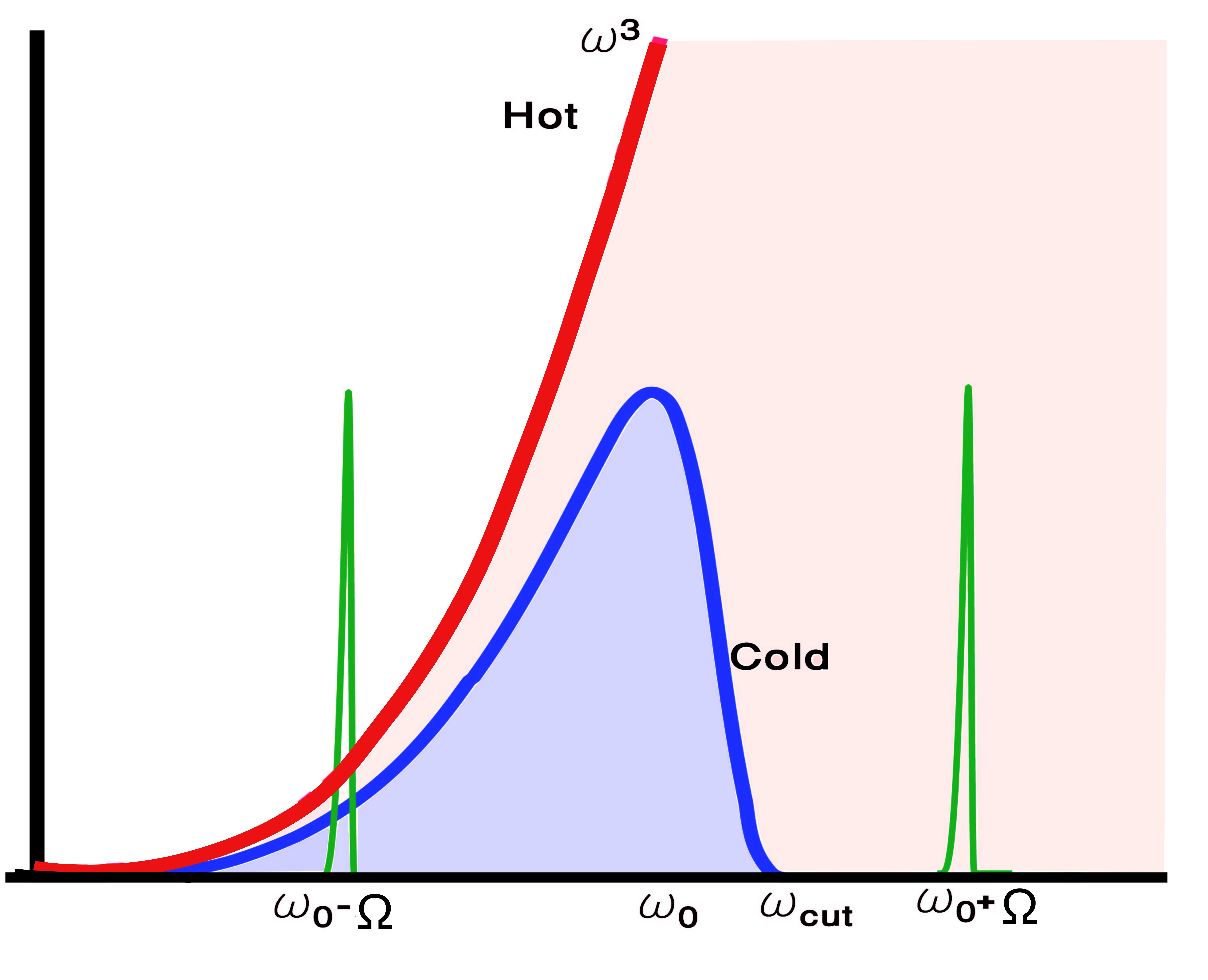}\protect\caption{\label{fig:16}Separation of the hot and cold bath spectra required
for the universal heat machine in figs. \ref{fig:14}-\ref{fig:15}.
The hot-bath spectrum $G^{h}(\omega)$ (red) is that of black body
radiation, rising with mode density as $\omega^{3}$, whereas the
cold bath spectrum $G^{c}(\omega)$ (blue, Lorentzian or Ohmic) extends
up to $\omega_{cut}$. The periodic modulation ($\pi$-flips) of the
qubit creates two sidebands (green): the lower overlaps both bath
spectra, but the upper only overlaps the hot-bath spectrum \cite{gelbwaser2013minimal}.}
\end{figure}
where $q=\pm1,\pm2,...$ is the harmonic index, $a=c,h$ is the bath
index, and $\mathcal{L}_{q}^{a}\tilde{\rho_{S}}$ describes the steady-state
with the qubit resonance $\omega_{0}$ shifted by $q\Omega$, $\Omega$
being the periodic modulation frequency. This decomposition of the
Liouvillian effectively replaces each bath by multiple $q$-harmonic
``sub-baths'' with differently shifted coupling spectra. The merit
of this equation is that it employs Lindblad (completely-positive)
dynamics but still allows for non-flat bath spectra. This steady-state
expansion can serve to evaluate the heat currents exchanged among
the multiple harmonic ``sub-baths'' via the qubit. These currents
can be controlled and optimized by the modulation and the bath-spectrum
engineering. Their signs will determine whether the machine functions
as QHE or QR, \emph{without the need} for traditional stroke cycles
(Carnot, Curzon, Otto). We have studied these QHE and QR models in
optomechanical \cite{2015NatSR...5E7809G} and spin-ensemble \cite{2014PhRvE..90b2102G}
setups.

We contend that such a periodically-modulated control qubit coupled
to both baths is a \textit{minimal model} for QHE and QR (Fig. \ref{fig:15}).
It is remarkable that such a simple model allows for both QHE and
QR actions, by contrast to previous models that required 3-level \cite{geva1992classical,kosloff2013quantum}
or coupled-qubit \cite{Gisin_Quantum_cryptog_2002,Sergienko_2005_QCC}
configurations. 

\begin{figure}
\begin{centering}
\includegraphics[width=0.5\textwidth]{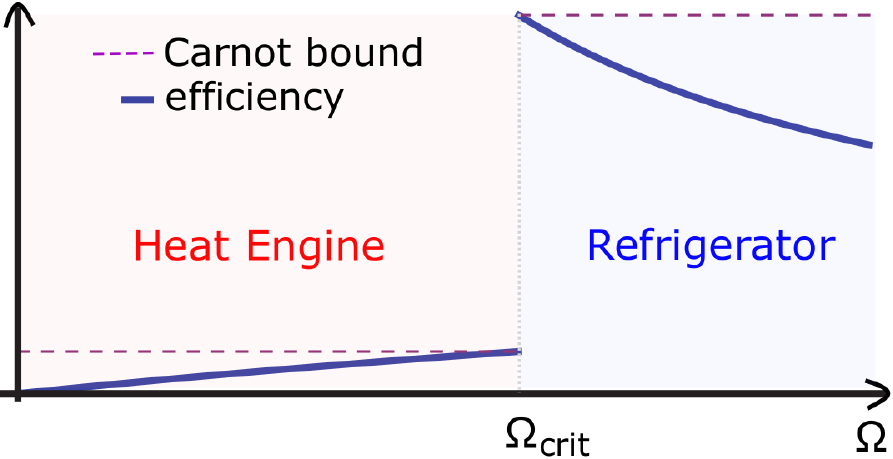}
\par\end{centering}

\protect\caption{\label{fig:17}Performance of the universal heat machine depicted
in figs. \ref{fig:14}-\ref{fig:16} (under the assumption that the
spectra of the two baths are non-overlaping (separated)). The efficiency
is plotted as a function of the modulation rate $\Omega$ \cite{gelbwaser2013minimal}.
The efficiency in the heat engine regime rises with $\Omega$ up to
the Carnot bound at $\Omega=\Omega_{crit}$. At higher $\Omega$ the
machine switches over to the refrigerator regime with a coefficient
of performance that decreases from the Carnot bound.}
\end{figure}

Under $\pi$-flip (periodic) modulation of the qubit there are only
two dominant Floquet harmonics at $\omega=\omega_{0}\pm\Omega$. If
the cold-bath coupling spectrum $G^{c}(\omega)$ has an upper cutoff,
such that the hot-bath spectra $G^{h}(\omega=\omega_{0}+\Omega)$
dominates at high energies, but $G^{c}(\omega=\omega_{0}-\Omega)$
dominates at low frequencies (Fig. \ref{fig:16}), then the heat flow
from the cold to the hot bath (cold current) $J_{c}$ is proportional
to the product of the respective hot- and cold-bath spectra, 
\begin{equation}
J_{c}\propto G^{h}(\omega_{0}+\Omega)G^{c}(\omega_{0}-\Omega).
\end{equation}
Under this condition $J_{c}$ is positive if 
\begin{equation}
n_{h}(\omega_{0}+\Omega)<n_{c}(\omega_{0}-\Omega)
\end{equation}
where $n_{h}$ and $n_{c}$ are the respective thermal bath occupancies.
We then obtain QR action, namely, heat removal from the cold bath
and its dumping into the hot bath. The opposite inequality sign implies
work production. Thus we have a \emph{single control parameter}, the
modulation rate, $\Omega$: for low rates we have an engine (QHE)
and for high ones a heat pump (QR) (Fig.~\ref{fig:17}). However,
neither the QHE nor the QR action will happen if the bath spectra
are inappropriate, so that bath-spectrum engineering is crucial.

\subsection{Third-Law Issues\label{sub:6.3_Third-Law-Issues}}

The limits on cooling power and efficiency of this unconventional
QR can serve to probe the validity of one of the formulations of the
Third Law of Thermodynamics by Nernst that prohibits the attainability
of zero temperature in finite time. To this end, we explore the use
of a \textit{single driven qubit} \textit{simultaneously} coupled
to hot and cold spectrally non-flat baths. The possibility of cooling
a bath down to arbitrarily low temperatures, \emph{i.e.} \textit{cooling
rate scaling with temperature,} is thus a fundamental issue that reflects
on the applicability of the Third Law.. As a typical example we assume
that the QR pumps heat into an infinite hot bath, and out of a cold
bath whose \textit{heat capacity is finite} $c_{V}<\infty$, resulting
in $T_{c}=T_{c}(t)$. Strikingly, we may show that \emph{arbitrarily
low} \emph{temperature} may be reached at \textit{finite time} (non-exponentially
fast) by the heat pump, for an appropriate (magnon) cold-bath spectrum
thereby challenging the Third Law \cite{kolavr2012quantum}.

\section{Conclusions and Outlook\label{sec:7_Conclusions-and-Outlook}}

Progress in technologies such as quantum information processing (QIP)
and quantum precision measurements (QPM) or metrology is currently
restricted by our ability to either minimize the environment effects
or actively suppress them by \textquotedblleft dynamical decoupling\textquotedblright .
Based on our theoretical and experimental work, we advocate instead
taking advantage of the environment (bath) as a resource for quantum
technologies, provided that we optimize its beneficial effects, preferably
by non-unitary open-system manipulations that are less restrictive
and more robust than unitary operations. 

To this end, we have identified the following generic tools:
\begin{itemize}
\item \textbf{Universal dynamical control} of open quantum systems\textbf{,}
be it coherent or projective (nonunitary) that can be optimized (in
terms of energy investment) for the desired operational task and bath
spectrum at hand. Such optimized control may conform to one of two
paradigms: the quantum Zeno effect (QZE) or the anti-Zeno effect (AZE),
depending on the task (Sec. \ref{sec: 2_Control-within-the}). In
particular, we have introduced the use of dynamical control as a means
of maximizing the quantum Fisher information (or estimation accuracy)
regarding the bath spectrum (Secs. \ref{sub:2.1_Control-for-bath}-\ref{sub:2.2_Maximized-information-on}
). We have also resorted to our universal dynamical control for optimizing
the tradeoff between the fidelity and duration of quantum information
transfer via noisy and random media which we model as baths (Sec.
\ref{sub:2.3_Bath-mediated-transfer-of}).
\item \textbf{Bath geometry control} by mode confinement to one dimension
and spectral-cutoff design has been shown (Sec. \ref{sec:4_Long-range-bath-induced-dispersi})
to allow for drastic increase in the range and fidelity of bath-induced
atom-atom entanglement (Secs. \ref{sec: 3_Bath-induced-entanglement},
\ref{sub:4.1_Long-range-deterministic-entangl}) and, even more dramatically,
the giant enhancement of dispersion (Casimir) forces (Sec. \ref{sub:4.3_Long-range vacuum-forces}).
These dispersive mechanisms rely on virtual-quanta exchange via the
bath, which is enhanced by the engineered bath geometry, as opposed
to dissipative (real-quanta) exchange, which is suppressed by the
chosen bath geometry.
\item \textbf{Bath-spectra engineering} has been shown to be a prerequisite
for thermodynamic control: AZE employed for high-speed qubit cooling
(Sec. \ref{sec:5_Thermodynamic-control-via}); the intriguing possibility
of exceeding the Szilard-Landauer bound by taking advantage of system-bath
correlations (Sec. \ref{sub:6.1_Work-information-tradeoff-in}); the
operation of simple (minimal) quantum heat machines based on a periodically-modulated
qubit that can attain both high efficiency (near the Carnot bound)
and power (Sec. \ref{sub:6.2_Minimal-quantum-heat}); as well as challenging
Nernst\textquoteright s formulation of the Third Law for the cooling
of a magnon bath towards the absolute zero (Sec. \ref{sub:6.3_Third-Law-Issues}).
\end{itemize}
We are confident that, however intriguing the above results are, they
just barely \textquotedblleft scratch the surface\textquotedblright{}
insofar as bath-assisted quantum processes are concerned. Inevitably,
such processes are within the realm of quantum thermodynamics. In
order to be able to benefit from the quantum control tools discussed
above, we should revisit the foundations of thermodynamics and reformulate
its key concepts and laws by \emph{(i)} removing the system-bath partition;
\emph{(ii)} exploring coherence and entanglement effects on thermodynamic
variables and \emph{(iii)} substantiating such fundamental effects
by studies of different realizations: NV-center spins coupled to a
spin bath, spin-boson models and boson-boson models in optomechanics.

The corresponding conceptual goals of such future research are foreseen
to be as follows:
\begin{description}
\item [{a)}] Use bath engineering, \emph{i.e.} control of its dimensionality,
coupling spectrum and quantum state, as a key resource in an effort
to push the thermodynamic limits of quantum device performance: i)
long-time non-Markovian behavior is expected for qubits near-resonant
with an abrupt spectral cutoff of a quasi 1D bath previously studied
by us \cite{kof94}. Extensions of these effects to finite-temperature
baths may be prerequisites to pushing thermodynamics into the hitherto
unexplored strong-coupling regime where system-bath separability breaks
down. This regime is expected to allow for (partial) reversibility
of the entropy and work and thereby alter quantum heat engine (QHE),
quantum refrigerator (QR) and quantum memory device (QMD) performance.
ii) Novel methods of controlling system-bath coupling by measurements
or phase flips at intervals that violate Markovianity, can be developed
so as to steer the system-bath dynamics towards desired outcomes.
\item [{b)}] Reexamine the work-effciency Carnot limit derived within the
system-bath separability (weak-coupling) paradigm: Little is known
about the strong-coupling regime in thermodynamics, and we may not
rule out that it has surprises in store insofar as performance bounds
of quantum heat machines are concerned \cite{scully2003extracting},
since the known definitions of heat currents and power output no longer
apply in that regime. 
\item [{c)}] Discover quantum-operations speed limits in the thermodynamic
limit: the rates (speed) of quantum information storage and retrieval,
cooling and heat engine cycles of quantum systems coupled to thermal
baths have unknown thermodynamic bounds. To understand these bounds,
we should extend our previous studies of the third law \cite{kolavr2012quantum}
by discovering the scaling of bath cooling-rate with temperature T
as $T\rightarrow0$.
\end{description}
To conclude, thermal baths are promising to be ``more friends than
foes'' for exploiting the quantumness of systems that couple to such
baths, provided that appropriate dynamical control and bath engineering
are implemented.

\ack{}{We acknowledge G. A. {\'A}lvarez and D. Gelbwaser for useful discussions. We acknowledge the support of ISF, Bikura, BSF and MOST. }

\bigskip{}

\bibliographystyle{iopart-num}
\bibliography{Bibl_GK-AZ}

\end{document}